\documentclass[aps, prd, superscriptaddress,
reprint,amsmath,amssymb, 
]{revtex4-2}

\makeatletter
\def\@hangfrom@section#1#2#3{\@hangfrom{#1#2}#3}
\def\@hangfroms@section#1#2{#1#2}
\makeatother

\usepackage{mathtools}
\usepackage{xcolor}
\usepackage{graphicx}
\usepackage[T1]{fontenc}
\usepackage[utf8]{inputenc}
\usepackage[
separate-uncertainty=true,
list-units=single,
range-units=single,
retain-explicit-plus,
group-minimum-digits=3,
]{siunitx}
\DeclareSIUnit\angstrom{\protect \text {Å}}

\usepackage[
a4paper, 
textwidth=18cm, 
top=1.8cm,
bottom=2.5cm,
]{geometry}

\usepackage{newfloat}
\DeclareFloatingEnvironment[
    fileext=lof,
    name={Extended Data Fig.},
    listname={List of Extended Data Figures}
]{extfigure}
\DeclareFloatingEnvironment[
    fileext=lof,
    name={Fig.},
]{nfigure}
\DeclareFloatingEnvironment[
    fileext=lot,
    name=Extended Data Table,
    listname={List of Extended Data Tables}
]{exttable}
\DeclareFloatingEnvironment[
    fileext=lot,
    name={Table},
]{ntable}

\usepackage[scaled]{helvet}
\usepackage[eulergreek,EULERGREEK]{sansmath}
\usepackage[colorlinks, allcolors=blue]{hyperref}
\usepackage[all]{hypcap}
\usepackage{colortbl,xcolor}

\newcommand{\rev}[1]{\textcolor{black}{#1}}

\begin{document}


\title{
Microstructural Insights into Fast Ion Transport in Solid Electrolytes via Multiscale Modeling
}

\author{Yongliang Ou}
\email{yongliang.ou@imw.uni-stuttgart.de}
\affiliation{Institute for Materials Science, University of Stuttgart, Stuttgart, Germany}
\affiliation{Department of Materials Science and Engineering, Massachusetts Institute of Technology, Cambridge, MA, USA}

\author{Lena Scholz}
\email{scholz@mib.uni-stuttgart.de}
\affiliation{Institute of Applied Mechanics, University of Stuttgart, Stuttgart, Germany}

\author{Sanath Keshav}
\affiliation{Institute of Applied Mechanics, University of Stuttgart, Stuttgart, Germany}

\author{Yuji Ikeda}
\affiliation{Institute for Materials Science, University of Stuttgart, Stuttgart, Germany}

\author{Marvin Kraft}
\affiliation{Institute of Inorganic and Analytical Chemistry, University of Münster, Münster, Germany}
\affiliation{Institute of Energy Materials and Devices (IMD), IMD-4:~Helmholtz-Institut Münster, Forschungszentrum Jülich, Münster, Germany}

\author{Sergiy Divinski}
\affiliation{Institute of Materials Physics, University of Münster, Münster, Germany}

\author{Rafael Gómez-Bombarelli}
\affiliation{Department of Materials Science and Engineering, Massachusetts Institute of Technology, Cambridge, MA, USA}

\author{Wolfgang G. Zeier}
\affiliation{Institute of Inorganic and Analytical Chemistry, University of Münster, Münster, Germany}
\affiliation{Institute of Energy Materials and Devices (IMD), IMD-4:~Helmholtz-Institut Münster, Forschungszentrum Jülich, Münster, Germany}

\author{Felix Fritzen}
\email{fritzen@mib.uni-stuttgart.de}
\affiliation{Institute of Applied Mechanics, University of Stuttgart, Stuttgart, Germany}

\author{Blazej Grabowski}
\email{blazej.grabowski@imw.uni-stuttgart.de}
\affiliation{Institute for Materials Science, University of Stuttgart, Stuttgart, Germany}

\date{\today}

\begin{abstract}

Improving solid electrolytes is critical for high-performance all-solid-state batteries, yet the microstructural features that enable fast ion transport remain poorly understood. 
Here, we use multiscale modeling to resolve polycrystalline ion transport from atomic-scale hopping at grain boundaries to continuum-scale percolation, thereby providing insights into realistic solid-electrolyte microstructures. 
Accurate lightweight machine-learning potentials---developed via closed-loop active learning for exemplar argyrodites Li$_6$PS$_5X$, $X \in \{\mathrm{Cl}, \mathrm{Br}, \mathrm{I}\}$---are employed to integrate molecular dynamics with finite element simulations.
We find that diffusion barriers of the anion-ordered bulk scale linearly with anion radius. 
Grain boundaries exert opposite effects depending on the bulk: enhancing ion diffusion in low-diffusivity phases but suppressing it in fast-diffusing ones.
Li$_6$PS$_5$I exhibits non-Arrhenius transport behavior consistent with experimental observations. 
Our results clarify the pivotal role of grain boundaries in ion transport and guide a priori microstructural design of advanced solid electrolytes. 

\end{abstract}

\maketitle

\section*{Introduction}

The rising demand for all-solid-state batteries has placed solid-state electrolytes at the forefront of material research, with most efforts historically focused on optimizing superionic bulk conductivity~\cite{Jun2024Dec, Wang_He_Yang_Cai_Wang_Lacivita_Kim_Ouyang_Ceder_2023, Wang_Richards_Ong_Miara_Kim_Mo_Ceder_2015}. 
However, practical solid electrolytes are polycrystalline, exhibiting microstructures with dense grain-boundary (GB) networks~\cite{Reynaud2023May}.
Macroscopic conductivity arises from long-range ion transport that inevitably involves both grain interiors and extensive GB pathways~\cite{Dutra_Goldmann_Islam_Dawson_2025}.
Microstructure thus represents a critical---yet often underutilized---lever for tailoring ion transport.
Although GB effects in oxide electrolytes have been partially elucidated~\cite{Cojocaru-Mirédin_Schmieg_Müller_Weber_Ivers-Tiffée_Gerthsen_2022, Huang2020Dec, Murugan2007Nov}, the microstructural features that enable fast ion transport remain poorly understood.  
Microstructural insights are particularly limited in sulfide electrolytes, where the role of GBs is still unclear due to the experimental difficulty of characterizing their morphology and separating their contributions to conduction.

Computer simulations can provide insights into complex microstructures by quantitatively predicting the microstructure--conductivity relation.
While prior electronic-structure modeling has yielded valuable insights into bulk ion mobility~\cite{Jun2024Dec, Wang_Fu_Liu_Saravanan_Luo_Deng_Sham_Sun_Mo_2023, Wang_Richards_Ong_Miara_Kim_Mo_Ceder_2015}, modeling ion transport in polycrystalline solid electrolytes is far more difficult. 
The transport process spans atomic-scale hopping and continuum-scale percolation, both influenced by structural and thermal variations~\cite{Heo2021Dec}, highlighting the need for a multiscale modeling framework~\cite{Feng_Wang_Kim_Wan_Wood_Heo_2024}. 
Molecular dynamics (MD) simulations are widely used to investigate atomic-scale diffusion mechanisms, but applying them to polycrystals poses several challenges. 
First, interatomic potentials must describe complex ionic interactions at GBs. 
Second, large simulation cells comprising millions of atoms are required to represent the structural disorder inherent to GBs in polycrystals. 
Third, long simulation times are needed to resolve the equilibrated GB structures (unknown a priori), and to obtain statistically converged diffusivities. 
Classical potentials generally lack the fidelity to describe intricate atomic interactions at GBs. 
While ab initio MD based on density-functional theory (DFT) provides high accuracy, its computational cost has restricted prior studies to idealized models with single GBs~\cite{Sadowski2024Nov, Quirk_Dawson_2023}. 
Recent machine-learning potentials offer a promising alternative, but parameterizing them to balance accuracy and computational cost is non-trivial~\cite{Wang_Guo_Gao_Wang_Zhang_Deng_Shi_Zhang_Zhong_2025, Jung_Srinivasan_Forslund_Grabowski_2023, Kim2022Jun}. 
Even with these potentials, MD simulations of solid electrolytes remain limited to nanometer length scales and nanosecond timescales~\cite{Jalem2023Dec,Kim2022Jun}. 
Finite element (FE) simulations provide a rigorous way for evaluating macroscopic properties of heterogeneous microstructures at the continuum scale. 
Nevertheless, resolving thin and anisotropic GBs in polycrystals and maintaining their applicability across the wide range of diffusion kinetics in solid electrolytes is difficult~\cite{Scholz_Ou_Grabowski_Fritzen_2025}. 
Finally, robust schemes for bridging simulations across scales remain scarce, with most previous studies addressing each scale independently (Suppl.~Table~3; Refs.~\cite{Scholz_Ou_Grabowski_Fritzen_2025, Heo2021Dec}). 

\begin{nfigure*}[!htbp]
    \capstart
    \centering
    \includegraphics{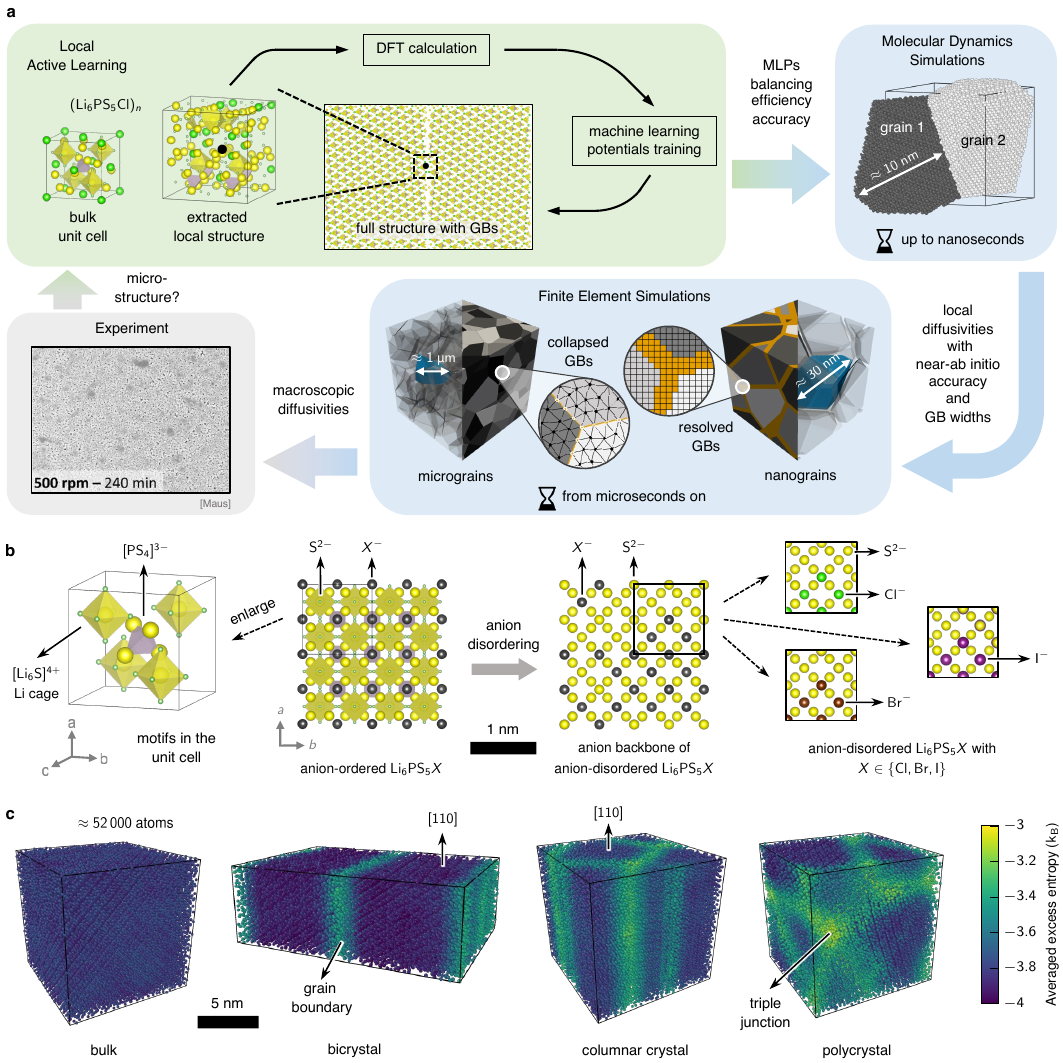}
    \caption{
    \textbf{\boldmath Multiscale modeling framework to resolve Li-ion transport and atomic and microstructural complexity of argyrodite solid electrolytes Li$_6$PS$_5X$ with $X \in \{\textrm{Cl}, \textrm{Br}, \textrm{I}\}$.} 
    \textbf{a}, 
    Lightweight machine-learning potentials (MLPs) balancing accuracy and efficiency are developed with closed-loop active learning, based on density-functional theory (DFT) calculations of local structures extracted from full microstructures with grain boundaries (GBs). 
    These potentials are applied in molecular dynamics (MD) simulations to obtain local Li-ion diffusivities. 
    Finite element (FE) simulations, parameterized with MD-derived diffusivities \rev{and GB widths}, are then performed on micrograins with collapsed GBs and on nanograins with resolved GBs using Fourier-accelerated nodal solvers, yielding macroscopic diffusivities. 
    The resulting predictions can be quantitatively compared with experimental measurements, and, when available, experimental microstructure data can inform end-to-end experiment--simulation studies. 
    A scanning electron micrograph of cathode composites containing processed solid electrolytes Li$_{5.5}$PS$_{4.5}$Cl$_{1.5}$ adapted from \citet{Maus_Lange_Frankenberg_Stainer_Faka_Schlautmann_Rosenbach_Jodlbauer_Schubert_Janek_et_al._2025} is shown as an example, in which GBs are poorly resolved.
    \textbf{b}, Atomic structures of bulk argyrodites featuring Li cages and ortho-thiophosphate ions. 
    Structural and chemical variations include anion disorder and halide substitution. 
    \textbf{c}, Atomistic models of microstructures including bulk phases, bicrystals with single-type GBs, columnar crystals, and polycrystals with random GBs. 
    The rotation axis is [110] for both bicrystals and columnar crystals.
    Triple junctions occur in columnar crystals and polycrystals, and both GBs and triple junctions exhibit increased excess entropy, indicating enhanced structural disorder. 
    }
    \label{fig:structure}
\end{nfigure*}

\begin{nfigure*}[!htbp]
    \capstart
    \centering
    \includegraphics{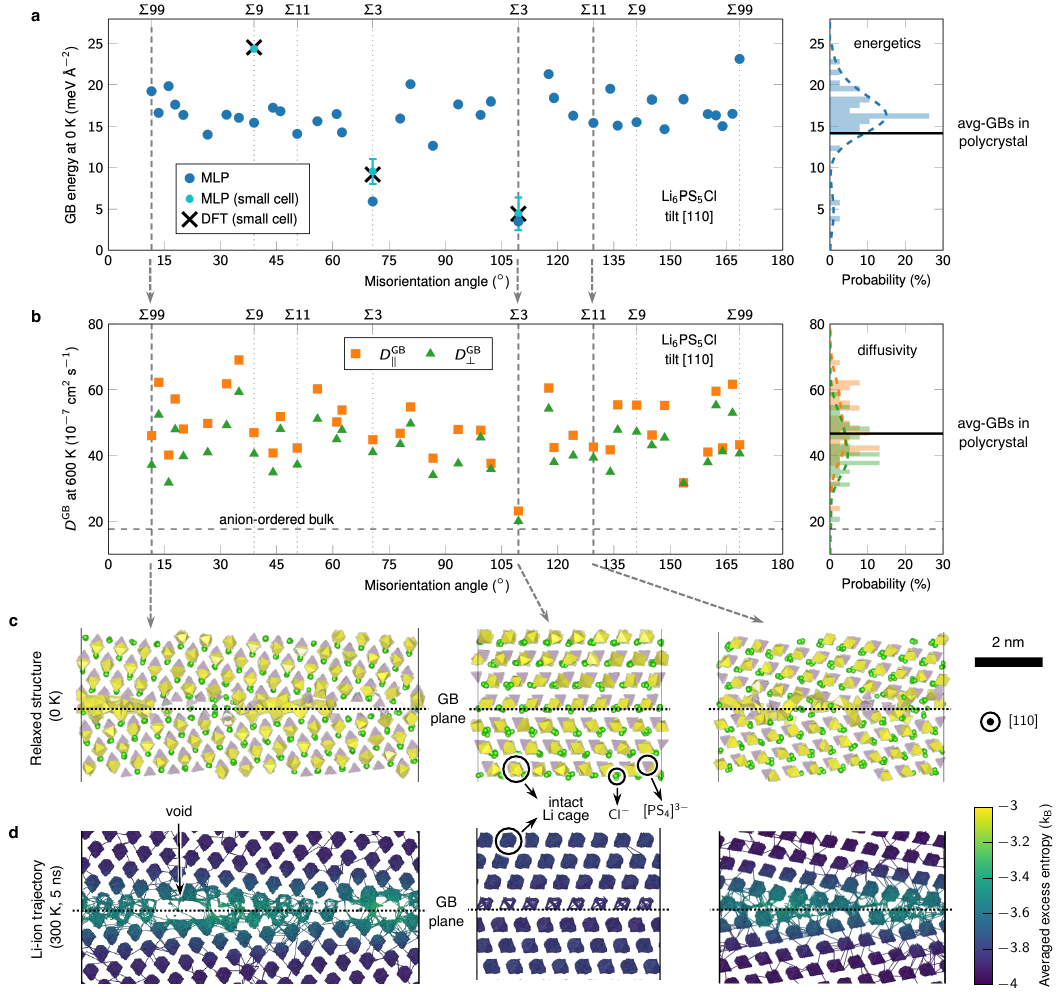}
    \caption{
    \textbf{\boldmath GB impacts in anion-ordered Li$_6$PS$_5$Cl.} 
    \textbf{a}, Predicted GB formation energy at \SI{0}{\kelvin}. 
    GB energies of $\Sigma3$ and $\Sigma9$ GBs in small simulation cells were calculated by density-functional theory (DFT) based on the structures relaxed by machine-learning potentials (MLPs), which validates the accuracy of MLPs.
    Error bars show the standard deviation across five independent potentials.
    \textbf{b}, Predicted GB self-diffusivities \smash{$D^\mathrm{GB}$} at \SI{600}{\kelvin}. 
    Diffusion within and across the GB planes is represented by \smash{$D^\mathrm{GB}_\parallel$} and \smash{$D^\mathrm{GB}_\perp$}, respectively. 
    Li-ion diffusivity is lower in the anion-ordered bulk than in GBs, and \smash{$D^\mathrm{GB}_\perp \approx 0.87 D^\mathrm{GB}_\parallel$}.
    Single-type tilt GBs were constructed with a rotation axis [110].
    Probability distributions of GB energy and diffusivity are shown, both exhibiting small variations in the GB configuration space, except for $\Sigma 3$ GBs. 
    GB energy and diffusivity averaged over all GBs within a polycrystalline model are consistent with those from single-type GBs. 
    \textbf{c}, Relaxed atomic structures of $\Sigma99$, $\Sigma3$, and $\Sigma11$ GBs at \SI{0}{\kelvin}. 
    Irregular yellow polygons at the $\Sigma99$ and $\Sigma11$ GBs highlight complex Li--S coordination.
    \textbf{d}, Li-ion trajectories from MD simulations at \SI{300}{\kelvin} over \SI{5}{\nano\second} for the three selected GBs. 
    In $\Sigma99$ and $\Sigma11$, GBs alter Li–S coordination of Li cages, producing structural disorder evidenced by increased excess entropy.
    Sub-nanometer voids appear in some GBs, such as $\Sigma99$. 
    }
    \label{fig:single}
\end{nfigure*}

Here, we propose a computational strategy to address the challenge of modeling ion transport in polycrystalline solid electrolytes (Fig.~\ref{fig:structure}\textbf{a}). 
We first develop a closed-loop active learning scheme based on DFT calculations of local atomic environments for parameterizing lightweight machine-learning potentials, specifically moment tensor potentials~\cite{Shapeev2016Sep}. 
These potentials are applied in large-scale MD simulations of a surrogate GB model with random GBs, capturing diffusion kinetics while accounting for thermodynamics and GB heterogeneity.
Then, we deploy the MD-derived self-diffusivities in FE simulations, modeling micrograins with collapsed GBs~\cite{Scholz_Ou_Grabowski_Fritzen_2025} and nanograins with resolved GBs using Fourier-accelerated nodal solvers~\cite{Leuschner2018a}. 
This approach provides a hierarchical resolution of Li-ion transport from the electronic-structure level to the device-relevant behavior, delivering quantitative predictions comparable with experimental measurements and enabling end-to-end experiment--simulation studies.

We apply the proposed strategy to exemplar sulfide solid electrolytes, Li$_6$PS$_5X$ argyrodites with $X \in \{\textrm{Cl}, \textrm{Br}, \textrm{I}\}$, which exhibit some of the highest known ionic conductivities~\cite{Deiseroth_Kong_Eckert_Vannahme_Reiner_Zaiß_Schlosser_2008} and are promising for applications. 
The morphologies and energetics of diverse GBs, including low-angle types, are systematically investigated. 
We evaluate GB impact on Li-ion transport in anion-disordered Li$_6$PS$_5$Cl and examine the effect of substituting Cl with Br or I. 
Integrating the resulting diffusivities into continuum-scale simulations quantifies the bulk-dependent grain-size effects in argyrodites. 
Overall, these findings provide insights for designing solid-electrolyte microstructures. 

\begin{ntable*}[!htbp]
    \capstart
    \caption{
    \textbf{Predicted Li-ion self-diffusivities and derived quantities of argyrodite solid electrolytes at 300~K.} 
    The anion disorder row indicates the degree of \textit{X}\!/S-anion disorder. 
    Diffusion coefficients \smash{$D^{\mathrm{GB}}$} are from MD simulations of polycrystals, representing averages over all GB structures within them. 
    The GB-to-bulk ratio quantifies the impact of GBs on diffusion in the polycrystal with $\uparrow$ and $\downarrow$ indicating increase and decrease, respectively. 
    The $f$ rows show the predicted characteristic Li-ion hopping frequencies, which are consistent with experimental values from~\citet{shotwell_tetrahedral_2025}. 
    \rev{Characteristic frequencies of the bulk and GBs are comparable for anion-disordered Li$_6$PS$_5$Cl and Li$_6$PS$_5$Br, but differ by about two orders of magnitude for anion-ordered Li$_6$PS$_5$I.} 
    }
    \label{tab:ratio}
    \vspace{0.1cm}
    \sffamily
        \sansmath
        \selectfont
\begin{ruledtabular}
        {
\begin{tabular}{lccccccc}
 System & \multicolumn{5}{>{\columncolor{gray!15}}c}{Li$_6$PS$_5$Cl} & Li$_6$PS$_5$Br  & Li$_6$PS$_5$I  \\
\hline \noalign{\vspace{2pt}}
Anion disorder (\%) & 0 &  25 & 50  & 75  & 100  & \rev{50} & 0 \\
$D^{\mathrm{bulk}}$ ($10^{-7}$\,cm$^{2}$s$^{-1}$) & 0.010 & 2.153 & 2.184 & 1.808 & 0.048 & \rev{1.522} & $6.2\times10^{-7}$ \\
$D^{\mathrm{GB}}$ ($10^{-7}$\,cm$^{2}$s$^{-1}$) & 0.092 & 0.401 & 0.959 & 0.690 & 0.136 & \rev{0.436} & $3.7\times10^{-5}$ \\
$D^{\mathrm{GB}} / D^{\mathrm{bulk}}$ & 8.79 $\uparrow$ & 0.19 $\downarrow$ & 0.44 $\downarrow$ & 0.38 $\downarrow$ & 2.85 $\uparrow$ & \rev{0.29 $\downarrow$} & 59.25 $\uparrow$ \\
$f^{\mathrm{bulk}}$ (Hz) & $8.61\times10^{6}$ & $1.77\times10^{9}$ & $1.80\times10^{9}$ & $1.49\times10^{9}$ & $3.92\times10^{7}$ & \rev{$1.25\times10^{9}$} & $5.05\times10^{2}$ \\
$f^{\mathrm{GB}}$ (Hz) & $7.57\times10^{7}$ & $3.30\times10^{8}$ & $7.90\times10^{8}$ & $5.68\times10^{8}$ & $1.12\times10^{8}$ & \rev{$3.57\times10^{8}$} & $2.99\times10^{4}$ \\ [3pt]
$f^{\mathrm{experiment}}$ (Hz) \tiny\textcolor{gray}{[Shotwell]} & \multicolumn{5}{>{\columncolor{gray!15}}c}{$2.50\times10^{7}$} & $1.37\times10^{7}$ & $1.27\times10^{5}$ \\
\end{tabular}       
        }
\end{ruledtabular}
\end{ntable*}

\section*{Results}

\subsection*{Energetics and diffusivity of GBs in anion-ordered argyrodites}

Argyrodite solid electrolytes exhibit a high degree of chemical complexity and compositional tunability. 
In fully ordered Li$_6$PS$_5X$, the bulk structure consists of Li-coordinated cages (Li$_6$S$^{4+}$) and PS$_4^{3-}$ tetrahedra (Fig.~\ref{fig:structure}\textbf{b}). 
The translationally immobile anion sublattice forms a backbone for Li$^+$ interstitial diffusion, and partial occupancy of S$^{2-}$ and $X^-$ introduces chemical disorder. 

Microstructure adds an extra layer of complexity in argyrodites. 
In GBs, structural and chemical disorder are coupled. 
To capture these effects, we investigate GBs embedded in three representative atomistic models (Fig.~\ref{fig:structure}\textbf{c}). 
Bicrystal models contain a single GB type, enabling isolated analysis of specific GB structures. 
Columnar crystals feature multiple GBs intersecting along a common rotation axis, whereas polycrystals form random GB networks. 
Both columnar and polycrystalline models include triple junctions in addition to GBs. 

Experimental characterization of microstructure in sulfides remains challenging, leaving GBs in synthesized samples poorly resolved (Fig.~\ref{fig:structure}\textbf{a}). 
In simulations, this implies that the type, distribution, and atomic structure of GBs are unknown a priori and must be determined in silico.
We construct bicrystal models of anion-ordered Li$_6$PS$_5$Cl with single-type GBs to probe GB morphology. 
GBs are labeled by the coincidence-site lattice index $\Sigma$, defined during model construction, indicating structural complexity.
Equilibrated GB structures, incorporating Li-ion disorder from diffusion and anion-sublattice distortions induced by GBs, are obtained using an MD--based annealing-and-quenching protocol~\cite{Ou2024Nov} (see Methods).

Figure~\ref{fig:single}\textbf{a} shows that the $\Sigma3$ GB at \SI{109}{\degree} has the lowest formation energy, \SI{\sim5}{\meV.\angstrom\textsuperscript{\textminus2}}, whereas other GBs, including low-angle GBs, range from \SIrange{12}{22}{\meV.\angstrom\textsuperscript{\textminus2}}.
Differences in formation energy stem from GB morphology, with each type imposing different structural modifications on the lattice.
In the $\Sigma3$ GB, Li-cage and PS$_4^{3-}$ environments are largely preserved, with rearrangements of these units and the halide ions, evident in the relaxed structure (Fig.~\ref{fig:single}\textbf{c}). 
By contrast, general GBs heavily distort Li cages and PS$_4^{3-}$ units, leading to complex Li--S coordination, anion-sublattice disorder, and lattice distortions. 
Forming GBs without altering Li-cage and PS$_4^{3-}$ units is expected to be rare, as GBs are two-dimensional defects distributed throughout the bulk.
These results underscore the importance of simulating realistic GBs beyond idealized cases such as $\Sigma3$ (cf.~previous studies in Suppl.~Table~3). 
The GB formation energies of Li$_6$PS$_5$Cl are low compared with oxide electrolytes (\SIrange{25}{56}{\meV.\angstrom\textsuperscript{\textminus2}}~\cite{Shiiba2018Sep}), and sulfide-based argyrodites are mechanically soft, both factors likely promoting the formation of abundant and random GBs in synthesized samples. 

\begin{nfigure*}[!htbp]
    \capstart
    \centering
    \includegraphics{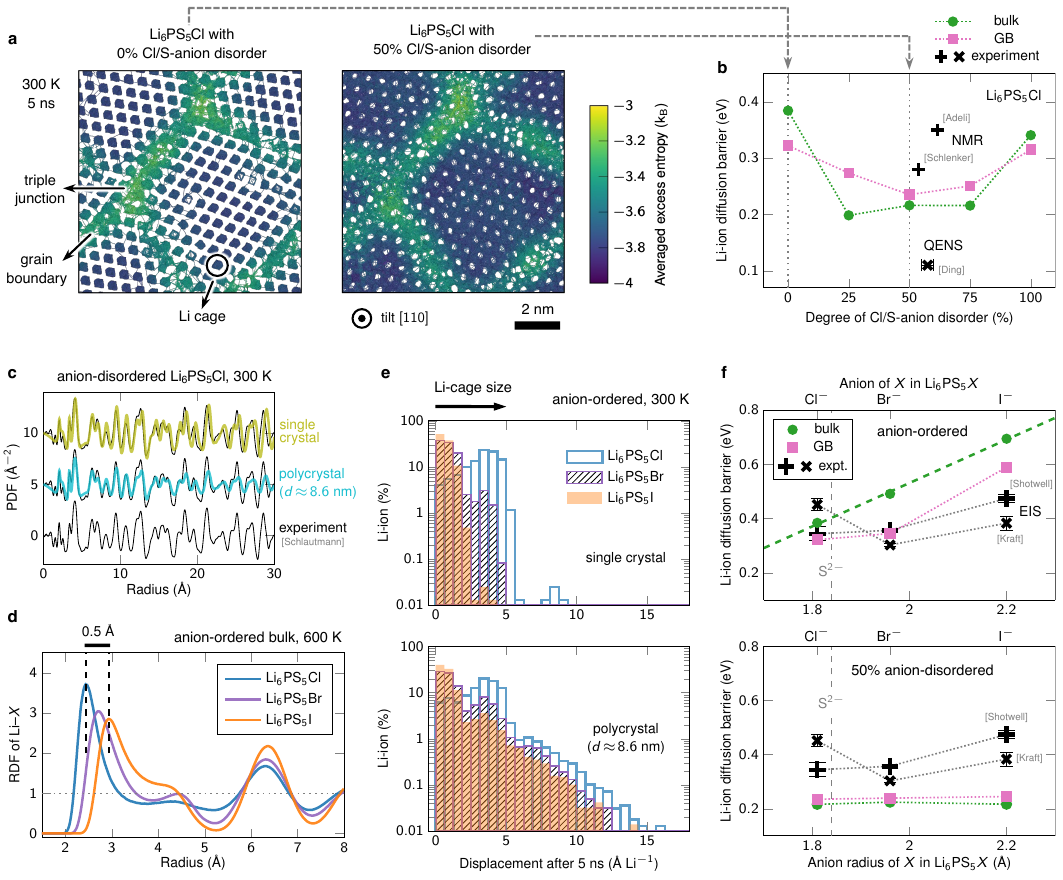}
    \caption{
    \textbf{GB effects in argyrodite solid electrolytes with anion disorder and halide substitution.} 
    \textbf{a},~Li-ion trajectories in the columnar polycrystal of Li$_6$PS$_5$Cl from MD at \SI{300}{\kelvin} over \SI{5}{\nano\second}. 
    GBs are characterized by increased excess entropy, indicating lattice distortions. 
    %
    \textbf{b},~Li-ion diffusion barriers in bulk and GBs (averages of all GB structures in the polycrystals) in Li$_6$PS$_5$Cl. 
    Measurements based on nuclear magnetic resonance (NMR; \citet{Adeli2019Jun, Schlenker2020Oct}) and quasielastic neutron scattering (QENS; \citet{Ding2025Jan}) show higher and lower barriers, respectively. 
    \textbf{c}, Comparison of simulated and experimental pair distribution functions (PDFs) for Li$_6$PS$_5$Cl. 
    Simulation results for single-crystalline ($y$-axis offset: \SI{+10}{\angstrom\textsuperscript{\textminus2}}) and polycrystalline (grain size $\sim$8.6~nm; \SI{+5}{\angstrom\textsuperscript{\textminus2}}) structures with 50\% anion disorder equilibrated at 300~K are shown.
    Experimental data from \citet{Schlautmann_Weiß_Maus_Ketter_Rana_Puls_Nickel_Gabbey_Hartnig_Bielefeld_et_al._2023} are shown with offsets for comparison.
    The simulated bulk PDF matches the experimental PDF, whereas the nanocrystalline PDF shows weaker peaks beyond 15~\AA, reflecting reduced structural coherence. 
    \textbf{d},~Time-averaged Li$^+$--halide radial distribution functions (RDFs) of bulk Li$_6$PS$_5X$ obtained from MD at \SI{600}{\kelvin}. 
    The first Li$^+$--halide coordination peak in Li$_6$PS$_5$I is attenuated and at a longer distance (\SI{\sim+0.5}{\angstrom}) compared to that in Li$_6$PS$_5$Cl. 
    Li$_6$PS$_5$Cl exhibits a more attenuated Li--$X$ RDF than others beyond \SI{4}{\angstrom}. 
    \textbf{e},~Distribution of Li-ion displacements in anion-ordered Li$_6$PS$_5X$ with $X \in \{\mathrm{Cl}, \mathrm{Br}, \mathrm{I}\}$ after MD at \SI{300}{\kelvin} over \SI{5}{\nano\second}. 
    Li$_6$PS$_5$Cl and Li$_6$PS$_5$Br exhibit intra-cage diffusion, whereas Li$_6$PS$_5$I shows primarily local hopping. 
    GBs enhance Li-ion diffusion in all polycrystalline argyrodites, and the increase is more pronounced in Li$_6$PS$_5$Cl, where more Li ions show inter-cage diffusion. 
    \textbf{f},~Decoupling anion ordering, anion species, and GB effects on Li-ion diffusion barriers.
    For anion-ordered argyrodites, the diffusion barrier increases linearly with anion radius in the bulk and GBs generally enhance diffusion.
    For anion-disordered argyrodites, the diffusion barrier remains low and shows no clear dependence on anion radius.
    The predicted trends agree with the experimental values and errors of~\citet{shotwell_tetrahedral_2025}, but differ from those of~\citet{Kraft2017Aug}; both were obtained via electrochemical impedance spectroscopy (EIS).
    }
    \label{fig:multiele}
\end{nfigure*}

We investigate diffusion in GBs using MD simulations of equilibrated bicrystal models, capturing Li-ion disorder and correlations, multiple diffusion pathways, GB-induced lattice distortions, and anion-sublattice vibrations. 
GB diffusivity and energetics follow a similar trend: The $\Sigma3$ GB at \SI{109}{\degree} has the lowest diffusivity, while others reach \SIrange{3e-6}{7e-6}{cm^2.s^{-1}} (Fig.~\ref{fig:single}\textbf{b}). 
At \SI{600}{\kelvin}, GB diffusivity is about threefold higher than in the bulk and slightly anisotropic, with faster Li-ion transport within the GB plane (Suppl.~Fig.~16). 
Li-ion trajectories at \SI{300}{\kelvin} (Fig.~\ref{fig:single}\textbf{d}) reveal that GB-induced lattice modifications create new low-barrier pathways, promoting inter-cage Li-ion hopping. 
Excess entropy (see Methods) quantifies the increased structural disorder at GBs arising from the anion sublattice and mobile Li ions. 
Subnanometer voids form at some GBs but have little effect on GB diffusivity. 

Each bicrystal represents a single GB configuration, providing a simplified model without triple junctions. 
To probe general configurations, we construct two-grain polycrystals embedded with random GB networks, serving as surrogate GB models (see Methods). 
Their averaged energetics and Li-ion diffusivities fall within the range of single-type GBs (histograms in Figs.~\ref{fig:single}\textbf{a} and \ref{fig:single}\textbf{b}). 
Columnar and polycrystalline models show comparable Li-ion diffusivity (Suppl.~Fig.~17). 
These results indicate that triple junctions have a minor effect in polycrystalline Li$_6$PS$_5$Cl.

\subsection*{GBs in argyrodites with anion disorder and substitution}

Our analysis has so far focused on anion-ordered Li$_6$PS$_5$Cl. 
In argyrodite solid electrolytes, anion chemistry---particularly disorder and substitution (Fig.~\ref{fig:structure}\textbf{b})---critically influences Li-ion transport. 
In the bulk lattice, site exchange between S$^{2-}$ and $X^-$ markedly promotes inter-cage Li$^+$ diffusion~\cite{Jeon_Ho_Cha_Chul_Jung_2024,Morgan_2021}.
Still, it remains unclear how GBs are coupled with anion chemistry and how this coupling influences Li$^+$ transport. 

To reveal GB effects in anion-disordered argyrodites, we employ Li$_6$PS$_5$Cl as the model system. 
Columnar and polycrystalline models are constructed from anion-disordered bulk, so their GB regions are influenced by the imposed disorder. 
Li-ion trajectories in the bulk from MD simulations at 300~K no longer exhibit distinct cages, indicating a homogeneous spatial distribution of diffusing Li ions (cf.~Fig.~\ref{fig:multiele}\textbf{a}). 
Anion disorder lowers the bulk diffusion barrier by \SI{\sim0.17}{eV} (0\% vs.~50\%; Fig.~\ref{fig:multiele}\textbf{b}), consistent with previous studies~\cite{Jeon_Ho_Cha_Chul_Jung_2024, Morgan_2021}. 
GBs increase the excess entropy in both anion-ordered and anion-disordered bulk phases, yet their influence on Li-ion diffusion differs. 
With ordered anions, the diffusion barrier at GBs is reduced by \SI{\sim80}{meV}, whereas with 50\% disorder, GBs increase the barrier by \SI{\sim10}{meV} relative to the bulk.

\begin{nfigure*}
    \capstart
    \centering
    \includegraphics{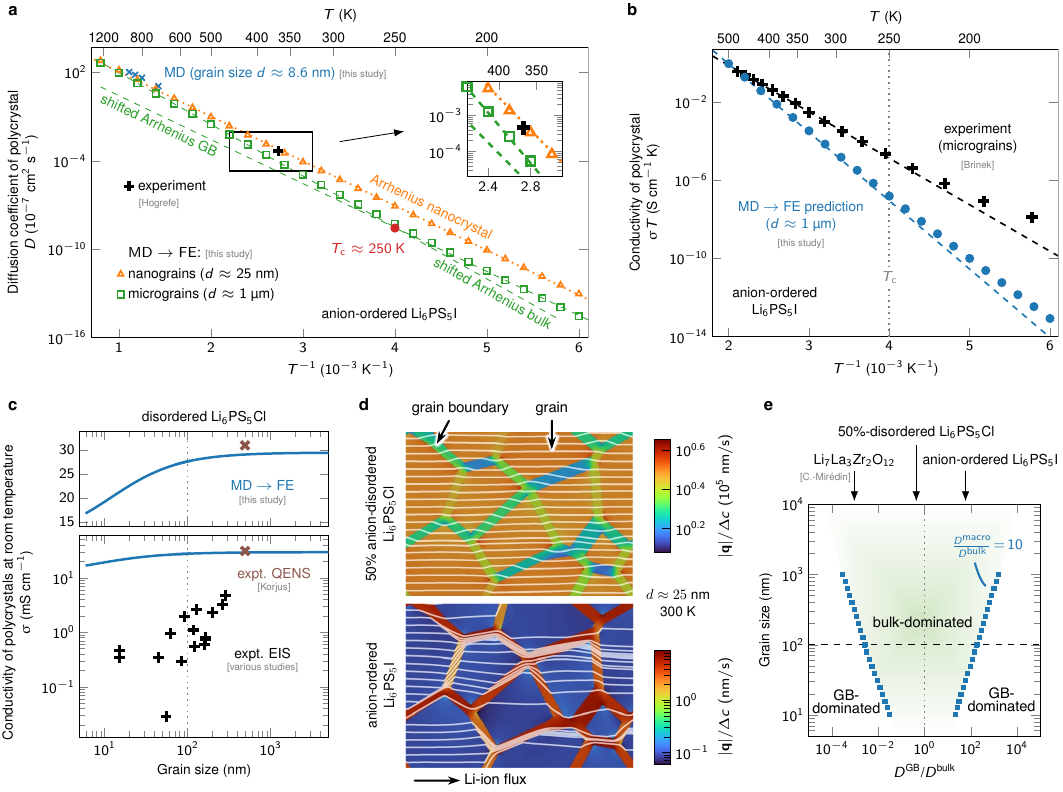}
    \caption{
    \textbf{Li-ion transport from atomic-scale diffusion to continuum-scale percolation.}
    \textbf{a}, Comparison of predicted and experimental diffusivities of anion-ordered Li$_6$PS$_5$I. 
    MD simulations are limited to small grain sizes and high temperatures, while the proposed approach (MD~$\rightarrow$~FE) provides accurate predictions for wide ranges of grain size and temperature. 
    Two diffusion regimes are identified from shifted Arrhenius fits of the calculated bulk and GB diffusivities, which also determine the transition temperature. 
    Predicted macroscopic diffusivities are in good agreement with nuclear magnetic resonance measurements (\citet{Hogrefe2021Oct}).   
    \textbf{b}, Non-Arrhenius transport behavior in anion-ordered polycrystalline Li$_6$PS$_5$I originating from GBs.
    Predicted and experimental transition temperatures ($T_{\mathrm{c}} \approx 250$~K; \citet{Brinek2020Jun}) and low-temperature activation energy changes (\qty{-0.10}{}~vs.~\qty{-0.12}{eV}) are in good agreement. 
    \textbf{c}, Grain-size effect on conductivities of anion-disordered Li$_6$PS$_5$Cl at room temperature. 
    Predictions for Li$_6$PS$_5$Cl with 50\% disorder show good agreement with quasielastic neutron scattering (QENS) measurements (\citet{Korjus_Mitra_Berrod_Vanpeene_Appel_Broche_Lyonnard_Villevieille_2025}), and exceed values from electrochemical impedance spectroscopy (EIS) measurements (\citet{Buchberger_Garbacz_Słupczyński_Brzezicki_Boczar_Czerwiński_2023, 
    Randrema_Barcha_Chakir_Viallet_Morcrette_2021, Arnold_Buchberger_Li_Sunkara_Druffel_Wang_2020, Yu_Ganapathy_Hageman_van_Eijck_van_Eck_Zhang_Schwietert_Basak_Kelder_Wagemaker_2018, Boulineau_Courty_Tarascon_Viallet_2012}). 
    \textbf{d}, Two-dimensional projections of the Li-ion flux $\mathbf{q}$ at 300~K for nanocrystalline (grain size $\sim$25~nm) Li$_6$PS$_5$Cl with 50\% disorder (\smash{$D^\mathrm{GB} < D^\mathrm{bulk}$}) and ordered Li$_6$PS$_5$I (\smash{$D^\mathrm{GB} > D^\mathrm{bulk}$}). 
    The fluxes are obtained from FE simulations under an applied horizontal concentration jump $\Delta c$. 
    \textbf{e}, Conditions under which GB effects substantially influence the diffusivity in polycrystalline solid electrolytes. 
    The cases where GBs alter the macroscopic diffusivity by more than an order of magnitude are referred to as GB-dominated, while bulk-dominated refers to the opposite. 
    The GB-to-bulk diffusivity ratios of 50\% anion-disordered Li$_6$PS$_5$Cl and anion-ordered Li$_6$PS$_5$I from the present study and of Li$_7$La$_3$Zr$_2$O$_{12}$ from \citet{Cojocaru-Mirédin_Schmieg_Müller_Weber_Ivers-Tiffée_Gerthsen_2022} at 300~K are indicated.
    }
    \label{fig:poly}
\end{nfigure*}

At \SI{300}{\kelvin}, GBs in anion-disordered Li$_6$PS$_5$Cl reduce Li-ion self-diffusivity by $\sim$60\% relative to the bulk yet retain high characteristic frequencies (\SI{\sim e8}{Hz}; Table~\ref{tab:ratio}), corroborating the measured low GB resistance~\cite{Ganapathy2019May} and indistinguishable GB contributions in impedance spectra~\cite{Kraft2017Aug}. 
Predicted bulk and GB diffusion barriers (Fig.~\ref{fig:multiele}\textbf{b}; \SI{\sim0.23}{eV}) and room-temperature diffusivities (Suppl.~Fig.~10; \SI{\sim e-7}{cm^2.s^{-1}}) fall within the range of experimental values obtained by different methods (\SIrange{0.11}{0.37}{eV}; \SIrange{3e-8}{7e-7}{cm^2.s^{-1}}). 
This suggests that GBs are not the primary source of the large experimental variations, which may instead arise from other defects~\cite{Faka2024Jan} or spatially inhomogeneous anion disorder.
Experimental pair distribution functions closely match predictions for bulk Li$_6$PS$_5$Cl (Fig.~\ref{fig:multiele}\textbf{c}), implying that the measured samples are dominated by micrograins with bulk-like local coordinations. 
Nanosizing argyrodites is predicted to weaken correlations beyond 15~\AA, providing an atomistic explanation for the pair distribution decay observed in experiments~\cite{Maus_Lange_Frankenberg_Stainer_Faka_Schlautmann_Rosenbach_Jodlbauer_Schubert_Janek_et_al._2025}.

We examine the effects of anions on Li-ion diffusion using polycrystalline models built from anion-ordered phases and varying the anions from Cl$^-$ to Br$^-$ or I$^-$. 
Radial distribution functions of the bulk, obtained from MD at \SI{600}{\kelvin}, are used to probe the coordination environment between Li-ion and anions (Fig.~\ref{fig:multiele}\textbf{d}). 
Compared with Li--Cl, the first peak of the Li--I radial distribution function is shifted outward by \SI{\sim0.5}{\angstrom} and attenuated. 
The increase in Li--$X$ bond length reflects the larger anion radius of I$^-$, while the first-peak softening may result from weaker interactions of Li--I vs.~Li--Cl. 
Apart from the first peak, other peaks in Li--I are sharper than in Li--Cl, indicating that the bulk lattice is more ordered in Li$_6$PS$_5$I than in Li$_6$PS$_5$Cl during MD simulations. 

Li-ion displacements in single crystalline anion-ordered phases after \SI{5}{\nano\second} of MD at \SI{300}{\kelvin} are analyzed (Fig.~\ref{fig:multiele}\textbf{e}). 
In the Li$_6$PS$_5$Cl bulk, most Li-ion displacements match the Li-cage size (\SI{\sim5}{\angstrom}), confirming intra-cage motion, whereas in Li$_6$PS$_5$I, displacements are mostly limited to \SI{\sim3}{\angstrom}, reflecting local hopping between adjacent Li-ion sites. 
Simulations predict that substituting Cl with I in bulk argyrodites increases the diffusion barrier by \SI{\sim0.3}{eV}.
Including Li$_6$PS$_5$Br reveals a clear linear scaling of Li-ion diffusion barrier with anion radius (Fig.~\ref{fig:multiele}\textbf{f}), with an increase of \SI{0.1}{\angstrom} leading to an \SI{\sim80}{meV} higher barrier.
A linear trend also holds for GB formation energy (Suppl.~Fig.~18). 
By contrast, this trend is not apparent in anion-disordered argyrodites.

Our systematic study enables decoupling the effects of anion ordering, anion species, and GBs.
In anion-ordered phases, GBs disrupt Li-ion cages in all Li$_6$PS$_5X$, leading to Li-ion displacements exceeding \SI{5}{\angstrom}.
These effects are anion-dependent. 
Diffusion is most enhanced at Br GBs, which reduce the diffusion barrier by nearly \SI{0.2}{eV}, compared with reductions of less than \SI{0.1}{eV} in the Cl and I systems.
For anion-ordered Li$_6$PS$_5$I, experimental values are more than \SI{0.1}{eV} lower than the predicted GB value, likely due to additional defects that further enhance Li-ion transport~\cite{shotwell_tetrahedral_2025,Faka2024Jan}. 
By contrast, in anion-disordered phases, GBs introduce a nearly uniform additional barrier of \SI{\sim0.02}{eV}.

\subsection*{Non-Arrhenius transport and grain-size effects}

Atomistic simulations provide mechanistic insights into Li-ion diffusion at GBs, but with limitations (cf.~MD results in Fig.~\ref{fig:poly}\textbf{a}). 
Simulations are limited to grain sizes of about \SI{10}{\nano\meter} and only a few GBs, making macroscopic properties of real materials inaccessible. 
Time scales are also limited to nanoseconds, much shorter than in experiments. 
To predict the macroscopic properties of polycrystalline solid electrolytes with micrometer-sized grains while retaining near-ab initio accuracy, we utilize the MD-derived diffusivities and GB width (Suppl.~Note~4) in continuum FE simulations for the three anion-ordered Li$_6$PS$_5X$ with $X \in \{\textrm{Cl}, \textrm{Br}, \textrm{I}\}$ and for Li$_6$PS$_5$Cl with varying ratios of anion disorder.

For Li$_6$PS$_5$I with micrograins (Fig.~\ref{fig:poly}\textbf{a}), the macroscopic diffusivity exhibits non-Arrhenius behavior, with a transition temperature around 250~K. 
Diffusion is dominated by the bulk at high temperatures, with a barrier near \SI{0.69}{eV}, and by GBs at low temperatures, with a barrier near \SI{0.59}{eV}. 
The transition temperature and the degree of non-Arrhenius behavior are halogen-dependent (Suppl.~Note~5). 
Reducing the grain size to the nanometer scale mitigates non-Arrhenius behavior. 
Simulation-predicted grain-size effects suggest that the experimentally measured diffusivity (\SI{4.6e-11}{cm^2\,s^{-1}}) corresponds to a polycrystal with an effective grain size of \SI{31}{nm} (Suppl.~Fig.~20). 
We convert the diffusivities of a polycrystal with grain sizes about \SI{1}{\micro\meter} into conductivities (see Methods; Fig.~\ref{fig:poly}\textbf{b}). 
Predictions from simulations agree with the experimentally observed non-Arrhenius transport behavior. 
Although the predicted conductivities are underestimated---likely due to an overestimation of the diffusion barrier for Li$_6$PS$_5$I (Fig.~\ref{fig:multiele}\textbf{f})---the transition temperature is well reproduced. 
This suggests that the non-Arrhenius behavior observed in experiments may originate from GBs in the samples. 

Experimentally, the grain-size effect in argyrodites remains unclear, as grain size is difficult to control and conductivity varies with synthesis (cf.~Fig.~\ref{fig:poly}\textbf{c}). 
We predict the macroscopic conductivity of Li$_6$PS$_5$Cl with 50\% anion disorder at 300~K. 
Conductivity approaches the bulk value as grain size increases, with the most significant changes occurring with nanograins. 
Good agreement is observed between predictions and quasielastic neutron scattering measurements.
In contrast, the predicted values are roughly an order of magnitude higher than the highest experimental values measured by electrochemical impedance spectroscopy, which probes macroscopic transport. 
This discrepancy is partially mitigated by considering voids in solid electrolytes (Suppl.~Fig.~22).
Additional discrepancy may arise from differences in density, pellet processing, or electrode--sample contact~\cite{Ohno_Bernges_Buchheim_Duchardt_Hatz_Kraft_Kwak_Santhosha_Liu_Minafra_et_al._2020}. 
Therefore, the predictions provide grain-size-dependent upper limits for the conductivities of Li$_6$PS$_5$Cl. 
To examine the impact of nanosized grains in solid electrolytes, Li-ion fluxes in disordered Li$_6$PS$_5$Cl and ordered Li$_6$PS$_5$I at 300~K are visualized using FE simulations (Fig.~\ref{fig:poly}\textbf{d}). 
In Li$_6$PS$_5$Cl, GBs show little effect on flux magnitude or orientation, whereas in Li$_6$PS$_5$I, GBs act as connecting channels for Li-ion transport, with Li-ion preferentially flowing through them. 
These results reflect the impact of GBs on Li-ion transport: They reduce diffusivity by half relative to the bulk in Li$_6$PS$_5$Cl, but enhance diffusivity by up to 60-fold in Li$_6$PS$_5$I (Table~\ref{tab:ratio}). 

The impact of GBs on macroscopic diffusivity or conductivity depends on both \smash{$D^{\mathrm{GB}}$} and grain size. 
We provide a precomputed map linking atomic-scale changes in diffusivity to continuum-scale effects based on FE simulations (Suppl.~Fig.~21). 
Using the criterion that GBs alter the total diffusivity by more than an order of magnitude, the bulk-dominated and GB-dominated regimes can be quantitatively estimated (Fig.~\ref{fig:poly}\textbf{e}). 
For solid electrolytes with micrograins ($d>\SI{100}{nm}$), the GB effect becomes significant when GB diffusivity is either two orders of magnitude higher or three orders of magnitude lower than the bulk.

\section*{Discussion}

Our simulations reveal that GB voids are subnanometer in size and that GBs do not significantly impede Li$^+$ diffusion in the Li$_6$PS$_5$Cl solid electrolyte.
The lower-than-predicted experimental conductivities likely stem from limited interparticle contacts in powder samples, even after processing~\cite{Ohno_Bernges_Buchheim_Duchardt_Hatz_Kraft_Kwak_Santhosha_Liu_Minafra_et_al._2020}.
Grain refinement, such as wet milling before mechanical pressing~\cite{Wang_Hao_Naik_Vishnugopi_Fincher_Yan_Raj_Celio_Yang_Fang_et_al._2024}, which reduces porosity and introduces GBs, can thus effectively enhance intergranular contact in sulfide solid electrolytes while preserving their superionic conductivity.

GBs in argyrodites exert markedly different effects depending on the underlying anion sublattice. 
In materials with fast bulk diffusion pathways, GBs act as barriers, consistent with the reported GB resistance in oxide-based solid electrolytes~\cite{Huang_Chen_Huang_Xu_Shao_Wang_Li_Wang_2020, Tenhaeff_Rangasamy_Wang_Sokolov_Wolfenstine_Sakamoto_Dudney_2014}. 
Reducing GB resistance through modification or minimizing GB density is essential to improve such materials~\cite{Huang_Chen_Huang_Xu_Shao_Wang_Li_Wang_2020}. 
In contrast, in systems with intrinsically low bulk diffusivity, GBs facilitate Li$^+$ transport, as exemplified by anion-ordered Li$_6$PS$_5$I.
Similar GB-induced diffusion enhancement has also been predicted in sulfides such as Na$_3$PS$_4$~\cite{Dawson_Canepa_Clarke_Famprikis_Ghosh_Islam_2019} and $\beta$-Li$_3$PS$_4$~\cite{Jalem2023Dec}, in the oxide LiZr$_2$(PO$_4$)$_3$~\cite{Nakano_Tanibata_Takeda_Kobayashi_Nakayama_Watanabe_2021}, and in the halide Li$_3$InCl$_6$~\cite{Quirk_Dawson_2023}.
Consequently, nanosizing can offer a strategy to improve potential solid electrolytes lacking intrinsic superionic conductivity.

Because experimentally controlling anion type and anion disorder independently is difficult, the impact of incorporating different anions in argyrodites on their intrinsic transport properties has been unclear~\cite{shotwell_tetrahedral_2025,Hanghofer2019}. 
Our results reveal a linear dependence between diffusion barrier and the anion radius in ordered argyrodites. A large radius mismatch between S$^{2-}$ and $X^{-}$ obstructs Li$^+$ pathways and increases the diffusion barrier, consistent with the observation for K$_3$O$X$~\cite{Zheng_Elgin_White_Wu_2025}. 
This trend contrasts with LISICON-type solid electrolytes, where increasing the cation radius lowers the diffusion barrier to as low as 0.1~eV~\cite{Bachman_Muy_Grimaud_Chang_Pour_Lux_Paschos_Maglia_Lupart_Lamp_et_al._2016}. 
Our finding indicates that lattice polarizability, tunable through halide substitution, may not be the dominant factor governing Li-ion diffusivity in argyrodites, contrary to earlier suggestions~\cite{Kraft2017Aug}.

Non-Arrhenius transport behavior has been reported for several solid electrolytes and attributed to local energy fluctuations within the bulk~\cite{Okada_Ikeda_Aniya_2015,Šalkus_Kazakevičius_Kežionis_Orliukas_Badot_Bohnke_2011}.
However, this explanation cannot hold for Li$_6$PS$_5$I, where the diffusion barrier decreases at low temperature~\cite{Hanghofer2019}, with a transition at 250~K---well above the structural transition observed experimentally at 160~K~\cite{Hanghofer2019, Kong2010_Argyrodites}. 
The simulations show that GBs can give rise to non-Arrhenius behavior in Li$_6$PS$_5$I, thereby providing a possible atomic-scale mechanism for the experimental observations~\cite{Hanghofer2019}. 
Our results for Li$_6$PS$_5X$ reveal that both the transition temperature and the magnitude of non-Arrhenius behavior are halogen-dependent (Suppl.~Note~5), consistent with the softened and ultimately absent anomaly observed for Li$_6$PS$_5$Br$_{0.5}$I$_{0.5}$ and Li$_6$PS$_5$Br, respectively~\cite{Hanghofer2019}. 
In halide-rich argyrodites, GBs hinder Li$^+$ transport, leading to a predicted increase in the low-temperature diffusion barrier, consistent with experiments~\cite{Adeli_Bazak_Huq_Goward_Nazar_2021}. 
Moreover, our results reveal that nanosizing suppresses the non-Arrhenius response of Li$_6$PS$_5$I, in line with experiments~\cite{Brinek2020Jun}. 
Because non-Arrhenius behavior near room temperature can destabilize battery performance, it can be practically mitigated by controlling grain size or tuning GB diffusivity.

In conclusion, multiscale modeling integrating molecular dynamics and finite element simulations is accurate and efficient for studying ion transport in polycrystalline solid-state electrolytes. 
This approach provides the atomic-to-macroscopic support for the experimentally observed non-Arrhenius Li$^+$ transport in argyrodites and offers insights for designing solid electrolytes in all-solid-state batteries. 
Grain boundaries (GBs) can enhance or suppress diffusion depending on bulk diffusivity, implying that nanosizing may activate transport in systems lacking intrinsic superionic behavior. 
For Li$_6$PS$_5$Cl, grain refinement can improve intergranular contacts without compromising conductivity. 
Our results also highlight the anion radius as a key compositional parameter. 
Looking forward, systematic screening of GB diffusivities can establish the basis for extending GB engineering to solid electrolyte interfaces. 


\onecolumngrid


\newpage
\twocolumngrid

\section*{Methods}

\subsection*{Multiscale modeling framework}

Machine-learning potentials for large-scale atomistic simulations should balance accuracy with computational cost. 
To obtain such potentials for argyrodite solid electrolytes, a closed-loop active learning scheme combined with systematic parameterization of moment tensor potentials (MTPs) is developed. 

First, given that structural and chemical complexities are concentrated in grain-boundary (GB) regions, local atomic environments are extracted from full structures in the active learning process  (Suppl.~Fig.~4\textbf{a}). 
The extraction of local structures substantially reduces the cost of density-functional theory (DFT) calculations. 
The preservation of stoichiometry of the extracted structure and the relaxation of atoms at the boundaries of the periodic cell are introduced to minimize artificial effects. 

Second, a quality-level--based scheme is employed to efficiently sample the configurational space (Suppl.~Fig.~4\textbf{b}). 
Initial potentials with a small number of fitting parameters are trained on short ab initio molecular dynamics (MD) trajectories and used to explore configuration space via active learning. 
The accuracy of the potential after active learning is evaluated, and if insufficient, a potential with more fitting parameters (more accurate and computationally intensive) is used for the next round of active learning. 
This iterative process continues with increasingly complex potentials, enabling systematic and data-efficient development of potentials for complex structures requiring ab initio accuracy. 

The final machine-learning potentials are optimized for both accuracy  (Suppl. Fig.~5) and efficiency (Suppl.~Fig.~6), guided directly by the target material properties rather than fitting or validation error.
We refer readers to Suppl.~Note~1 for further details. 

Finite element (FE) simulations are employed to extend the accessible timescales from nanoseconds to beyond microseconds and length scales from nanometers to polycrystalline systems with micrometer-sized grains (Suppl.~Fig.~4\textbf{c}). 
The statistically converged diffusivities obtained from large-scale MD simulations accelerated by the trained MTP, including the GB width parameters, serve as input for the continuum-scale FE simulations. 

Resolving GB layers in polycrystalline models presents computational challenges for mesh-based FE simulations. 
To address this, different approaches are implemented depending on grain size to balance the accuracy and computational efficiency. 
For nanocrystalline polycrystals, GBs are volume-resolved, and continuum diffusion equations are solved using fast Fourier transform (FFT)--based methods. 
For microcrystalline polycrystals, GB layers become relatively thin and are thus represented as thickness-collapsed finite elements. 
In both cases, diffusivity tensors of polycrystalline models are obtained through linear computational homogenization. 
Comparative tests reveal that the volume-resolved and thickness-collapsed GB representations complement each other, supporting their combined use as an effective strategy to address the challenges of length-scale separation and diverse kinetic regimes presented in polycrystalline solid electrolytes (Suppl.~Fig.~11). 
We refer readers to Suppl.~Note~3 for further details. 

The proposed multiscale modeling framework enables prediction of macroscopic diffusivities in polycrystalline solid electrolytes up to device-relevant scales (Suppl.~Fig.~4\textbf{d}). 
Its main strength lies in balancing accuracy and efficiency across the entire modeling process, through an optimized active learning scheme for machine-learning potentials, a data-efficient multiscale workflow, and tailored methods for resolving GBs in FE simulations. 
Moreover, our modeling framework can be readily extended to explicitly include additional microstructural features. 
For example, surface diffusivities obtained from atomistic simulations can be incorporated into continuum models to describe transport along pore surfaces, enabling multiscale modeling of porous polycrystalline solid electrolytes.

\subsection*{Active learning}\label{sec:doptimal}

In the MTP formalism~\cite{Podryabinkin2023Aug}, the total energy $E$ of a structure is the sum of the atomic energies $V(\mathfrak{n}_i)$ of each atom in the structure,
\begin{equation}
    E = \sum_{i=1}^{N}V(\mathfrak {n}_{i}),
\end{equation}
where $\mathfrak{n}_i$ is the local atomic environment of atom $i$, and $N$ is the number of atoms in the structure. 

Since the atomic energy given by an MTP has a non-linear dependence on its fitting parameters, the generalized D-optimality criterion and the MaxVol algorithm~\cite{Gubaev2018Jun, Gubaev2019Jan} based on the fitted MTPs are used for active learning (AL).
In AL, the term \emph{global-AL} refers to cases where uncertainty is quantified over the entire simulation model, whereas \emph{local-AL} refers to cases where uncertainty is quantified individually for each atom in the model. 
For global-AL, variables $B_{jk}$ are defined as the partial derivative of the total energy $E_{j}$ of an atomic structure $j$ predicted by the fitted MTP and the fitting parameters $\theta_{k}$ of the MTP:
\begin{equation}
    B_{jk} \coloneqq \frac{\partial E_{j}}{\partial \theta_{k}}, \qquad \textrm{for global-AL}. 
\end{equation}
In local-AL, variables $B_{jk}$ are defined as the partial derivative of the atomic energy $V({\mathfrak{n}_j})$ of a local atomic environment $\mathfrak{n}_j$ predicted by the fitted MTP and the fitting parameters $\theta_{k}$ of the MTP:
\begin{equation}
    B_{jk} \coloneqq \frac{\partial V(\mathfrak {n}_{j})}{\partial \theta_{k}}, \qquad \textrm{for local-AL}.
\end{equation}

Suppose that an MTP has $m$ fitting parameters and has been fitted to a training set, i.e., the fitting parameters are optimized or initialized.
For the MTP, an active set with $m$ elements ($m$ structures for global-AL or $m$ local atomic environments for local-AL) can be selected from the training set so that they maximize the absolute value of the determinant of the following square matrix:
\begin{equation}
    \mathbf{A} = 
    \begin{bmatrix}
        B_{11} & B_{12} & \cdots & B_{1m} \\
        B_{21} & B_{22} & \cdots & B_{2m} \\
        \vdots & \vdots & \ddots & \vdots \\
        B_{m1} & B_{m2} & \cdots & B_{mm} \\
    \end{bmatrix}.
\end{equation}
In the active set matrix $\mathbf{A}$, each row corresponds to an element, and each column corresponds to the partial derivative of a fitting parameter. 

For a specific structure or local atomic environment $x$, the structural descriptor vector of $x$ can be obtained based on a fitted MTP,
\begin{equation}
    \mathbf{B}_{x} =
    \left(
        B_{x1}, B_{x2} , \cdots , B_{xm}
    \right). 
\end{equation}
Together with the active set matrix $\mathbf{A}$ of the fitted MTP, the extrapolation grade~\cite{Podryabinkin2023Aug} of $x$ can be calculated as
\begin{equation}
    \gamma_x = \max | \mathbf{B}_{x} \mathbf{A}^{-1} |.
\end{equation}
It can be interpreted that, given the training set and the fitted MTP, $x$ is evaluated as interpolative if $\gamma_x < 1$ and as extrapolative if $\gamma_x > 1$~\cite{Podryabinkin2017Dec}. 

In the present study, MTPs with eight radial basis functions were used. 
The minimum and maximum cutoffs were set to \SI{1.5}{} and \SI{5}{\angstrom}, respectively. 
The fitting weights for the energies and atomic forces were set to $N_{\mathrm{at}}^{-1}$ and \SI{0.001}{\angstrom\textsuperscript{2}}, respectively, where $N_{\mathrm{at}}$ is the number of atoms in the corresponding cell. 
The local extrapolation grade thresholds for selecting and breaking were 1.4 and 5, respectively. 
MTPs trained up to level 18 were used for all simulation models in the production runs. 
For the same material system, the training and validation root-mean-squared errors of all utilized MTPs are below \SI{10}{meV\,atom^{-1}} in energy (Suppl.~Fig.~5). 
Details of the parameters used in AL are provided in Suppl.~Note~1 and Suppl.~Figs.~1--3.

\subsection*{Simulation models}

The bicrystal GB models were constructed based on the ideal anion-ordered bulk of Li$_6$PS$_5$Cl and coincidence-site lattice theory~\cite{Cheng2018Dec} following our previous study~\cite{Ou2024Nov}. 
The geometry of a GB structure can be labeled by the integer index $\Sigma$, which is defined as the ratio between the coincidence unit cell volume and the rotated unit cell volume~\cite{Brandon_1966}.
The rotation axis of the two grains in the bicrystal models was set along [110], while the rotation angle between them was varied to generate distinct GB configurations. 
Supercells were constructed to ensure that adjacent periodic GBs were separated by more than \SI{50}{\angstrom}. 
The grains were slightly shifted in the normal direction of the GB plane to ensure that atomic distances of all atom pairs are larger than \SI{1.8}{\angstrom}. 

The atomistic columnar crystal and polycrystalline models consist of two grains in cubic cells and were constructed using \textsc{atomsk}~\cite{Hirel2015Dec} with periodic boundary conditions applied. 
For the columnar crystal, the rotation axis is [110] and the misorientation angle is \SI{\sim61}{\degree}. 
The position and crystallographic orientations of the two grains in the polycrystalline models were randomized, with the resulting misorientation angles \SI{\sim44}{\degree}, \SI{6}{\degree}, and \SI{8}{\degree} for the rotation axes [100], [010], and [001], respectively. 
The resulting atomic models thus contain distinct, randomly oriented GBs, serving as representative surrogate GB models. 
For each grain in the columnar and polycrystalline models, the stoichiometry of elements was ensured by removing atoms at GB regions. 
To construct anion-disordered models, S atoms inside the Li cage and Cl atoms were randomly selected according to the disorder percentage and exchanged, following our previous study~\cite{Ou2024Nov}. 
All simulation cells maintain the stoichiometry of Li$_6$PS$_5X$, ensuring formal charge neutrality. 
The local entropy of atomistic models was calculated based on the two-body excess entropy~\cite{Piaggi_Parrinello_2017} implemented in \textsc{OVITO}~\cite{ovito}. 
Atomistic models were visualized with \textsc{OVITO} and \textsc{VESTA}~\cite{Momma_Izumi_2008}. 
Pair distribution functions were calculated using \textsc{DiffPy-CMI}~\cite{Juhás_Farrow_Yang_Knox_Billinge_2015}. 

Polycrystalline microstructures in the continuum simulations were generated using three-dimensional periodic Voronoi tessellations~\cite{Fritzen2009} within a unit-cube domain, providing a simplified representation of realistic GB networks~\cite{Chen2007}.
This approach is appropriate for the present study, as the GB diffusivity averaged over all GBs in the surrogate atomic model was used as input for the continuum simulations.
The approach remains justified even when distinct GB diffusivities are used in the FE simulations (Suppl.~Fig.~20\textbf{a}), as the variation in diffusivities across different GB types is small.
The same tessellations were employed for both volume-resolved and thickness-collapsed GB representations, ensuring that the two approaches remain directly comparable. 
The GB width was set to \SI{2.5}{\nano\meter}, consistent with MD-resolved atomic structures (Suppl.~Note~4).

In continuum simulations, different discretizations were applied for the two GB representations (Suppl.~Fig.~4\textbf{c}). 
To resolve the GB, a structured voxel grid of prescribed resolution was generated using \textsc{MSUtils} (see Code availability), and each voxel was assigned an MD-derived diffusivity according to its region in the polycrystalline model. 
In the collapsed approach, a GB-conforming second-order mesh was generated with \textsc{Neper}~\cite{Quey2022,Quey2011} and then enriched with the interface elements~\cite{Scholz_Ou_Grabowski_Fritzen_2025}. 
The GB thickness was treated as an implicit parameter of the constitutive model, without being explicitly represented in the discretization.

Continuum simulations in production runs were carried out on a nearly isotropic polycrystalline model containing 27 grains, considered a realistic representative volume element (RVE) for polycrystalline solid electrolytes. 
The consistency of the two GB representations was tested using an idealized two-grain polycrystalline model to reduce computational cost.
Symmetrically placing the two tessellation seeds produces diamond-shaped grains, and the resulting model possesses isotropic transport properties. 
Voxel-based images of the resolved geometry and the corresponding collapsed crinkle-cut geometry for both the realistic and idealized models are shown in Suppl.~Figs.~11\textbf{a} and~11\textbf{b}. 
The continuum simulation models were visualized using \textsc{ParaView}~\cite{AHRENS2005717}.
Schematic figures were created using Microsoft PowerPoint.

\subsection*{Ab initio and atomistic simulations}

Electronic-structure calculations were performed using DFT as implemented in \textsc{VASP}~\cite{Kresse1999Jan,Kresse1996Oct,Kresse1995Dec}. 
The projector augmented-wave method~\cite{Blochl1994Dec} and the generalized gradient approximation in the Perdew--Burke--Ernzerhof (PBE) parametrization~\cite{Perdew1996Oct} were used. 
Electrons in the atomic orbitals $1s^22s^1$, $3s^23p^3$, $3s^23p^4$, $3s^23p^5$, $4s^24p^5$, and $5s^25p^5$ were treated as valence electrons for Li, P, S, Cl, Br, and I, respectively. 
The plane-wave cutoff was set to \SI{500}{eV}, and the energy was converged to less than \SI{e-3}{eV} per simulation cell.  
Ab initio MD simulations were performed with the conventional unit cell of Li$_6$PS$_5X$ with $X\in\{\mathrm{Cl}, \mathrm{Br}, \mathrm{I}\}$, containing 52 atoms, at \SI{1500}{\kelvin} using the Nosé--Hoover thermostat and the canonical ensemble implemented in \textsc{VASP}. 
The time step and the Nosé mass were set to \SI{2}{fs} and \SI{3}{u.\angstrom\textsuperscript{2}}, respectively. 
The reciprocal space was sampled using the Gaussian smearing with a width of \SI{0.03}{eV} and a $\Gamma$-centered $2\times2\times2$ $\mathbf{k}$-point mesh (\SI{416}{kp.atom}). 
For electronic-structure calculations of the extracted structures with about 200 atoms, the reciprocal space was sampled at the $\Gamma$ point. 

Large-scale MD simulations were performed by \textsc{LAMMPS}~\cite{Thompson2022Feb} with the fitted MTPs, with a time step of \SI{2}{fs}. 
The energy of the relaxed structure was obtained by the MD--based annealing-and-quenching approach: The GB-containing atomic models were first equilibrated at~\SI{600}{\kelvin}, then cooled to near \SI{0}{\kelvin} over \SI{1}{\nano\second}, and finally relaxed to \SI{0}{\kelvin}~\cite{Ou2024Nov}. 
The GB energy $\Gamma$ was calculated by
\begin{equation}
    \Gamma = \frac{E_{\mathrm{GB}}-E_{\mathrm{bulk}}}{A_{\mathrm{tot}}},
\end{equation}
where $E_{\mathrm{GB}}$ and $E_{\mathrm{bulk}}$ are the energies of the relaxed models with GB and with only bulk, respectively, and $A_{\mathrm{tot}}$ is the total GB area in the simulation cell. 
For the MD simulations targeting diffusion properties, the structures were first equilibrated with the \textit{NPT} ensemble at zero pressure for \SI{0.2}{ns}. 
The self-diffusion coefficient $D$ at temperature $T$ was calculated by
\begin{equation}
    D(T) = \frac{\langle u^2\rangle_T}{2dt},
\end{equation}
where $\langle u^2\rangle_T$ is the mean square displacement of Li atoms in MD, and $d$ and $t$ are the number of dimensions and simulation time, respectively. 
The production runs were performed in the \textit{NVE} ensemble for \SI{5}{\nano\second}, and the simulation window from \SIrange{3}{5}{ns} was used to calculate diffusion coefficients. 

GB diffusion coefficients were extracted from the diffusion coefficients derived from the simulation model with GBs. 
A GB width of \SI{2.5}{\nano\meter} was used,  consistent with our previous study~\cite{Ou2024Nov}. 
Within the normal diffusion regime, an Arrhenius relation for diffusion coefficients can be defined as
\begin{equation}
    D(T) = D_{0}\exp\left(-\frac{E_{\mathrm{a}}}{\mathrm{k_B} T}\right),
\end{equation}
where $D_0$ is a constant, $E_{\mathrm{a}}$ is the diffusion barrier, and $k_{\mathrm{B}}$ is the Boltzmann constant. 

Ionic conductivity $\sigma$ is related to self-diffusivity $D$ via a modified Nernst--Einstein equation~\cite{Uitz_Epp_Bottke_Wilkening_2017},
\begin{equation}
    \sigma = \Lambda\frac{(z\mathrm{e})^2D}{\mathrm{k_B}T},
\end{equation}
where $z$ is the valence of the mobile ion ($z=1$ for Li$^+$), e is the elementary charge, $\mathrm{k_B}$ is the Boltzmann constant, and $\Lambda$ is a system-dependent coefficient related to the concentration of the mobile ions and the Haven ratio. 
For Li$_6$PS$_5$Cl, $\Lambda \approx \SI{2.181e-2}{Li\,\angstrom\textsuperscript{-3}}$ at \SI{298}{\kelvin} was used, following our previous study~\cite{Ou2024Nov}. 
The same $\Lambda$ was used to convert the diffusion coefficients obtained from quasielastic neutron scattering measurements in Ref.~\cite{Korjus_Mitra_Berrod_Vanpeene_Appel_Broche_Lyonnard_Villevieille_2025}. 
For Li$_6$PS$_5$I, the conductivity from Ref.~\cite{Brinek2020Jun} and the diffusivity at \SI{366}{\kelvin} from Ref.~\cite{Hogrefe2021Oct} were used, resulting in $\Lambda \approx \SI{1.590e-1}{Li\,\angstrom\textsuperscript{-3}}$. 

The residence time $\tau$ of inter-cage Li-ion jumps was estimated as
\begin{equation}
    \tau = \frac{l^2}{6D},
\end{equation}
where $l$ denotes the inter-cage jump distance. The corresponding characteristic frequency $f$ was then obtained as 
\begin{equation}
    f = \frac{1}{\tau}.
\end{equation}
The inter-cage jump distance of anion-disordered Li$_6$PS$_5$Cl was reported to be \SI{2.70}{\angstrom}~\cite{Ding2025Jan}. 
Given the similar radial distribution functions, the same value (\SI{2.70}{\angstrom}) was adopted for the anion-ordered phases and GBs.
The relaxed lattice constants of the anion-ordered Li$_6$PS$_5$Cl, Li$_6$PS$_5$Br, and Li$_6$PS$_5$I structures (cubic, $F\bar{4}3m$ symmetry) were calculated as \SI{10.28}{\angstrom}, \SI{10.31}{\angstrom}, and \SI{10.35}{\angstrom}, respectively.
Assuming linear scaling with lattice parameter, the corresponding jump distances for Li$_6$PS$_5$Br and Li$_6$PS$_5$I were approximated as \SI{2.71}{\angstrom} and \SI{2.72}{\angstrom}. 

\subsection*{Continuum simulations}

On the continuum scale, we considered an ion-concentration field $c(\mathbf{x})$ defined at every position $\mathbf{x}$ within the simulation domain. 
The relation between the ion-concentration field and the ion flux $\mathbf{q}$ was described by classical Fickian diffusion, 
\begin{align}
    \mathbf{q}(\mathbf{x}) = \begin{cases}
        - \mathbf{D}^{\mathrm{bulk}}\nabla c(\mathbf{x}) & \text{in bulk},\\
        - \mathbf{D}^{\mathrm{GB}}\nabla c(\mathbf{x}) & \text{in GB},
    \end{cases}
\end{align}
where \smash{$\mathbf{D}^{\mathrm{bulk}}$} and \smash{$\mathbf{D}^{\mathrm{GB}}$} are diffusion tensors of the bulk and GB, respectively. 
The bulk was assumed isotropic, i.e. \smash{$\mathbf{D}^{\mathrm{bulk}} = D^{\mathrm{bulk}}\mathbf{I}$}, with the scalar diffusivity \smash{$D^{\mathrm{bulk}}$} derived from atomistic simulations. 
GBs were defined as transversely isotropic: Diffusion within the GB plane was described by \smash{$D^{\mathrm{GB}}_\parallel$}, whereas diffusion perpendicular to the plane was governed by \smash{$D^{\mathrm{GB}}_\perp$}. 
Both \smash{$D^{\mathrm{GB}}_\parallel$}and \smash{$D^{\mathrm{GB}}_\perp$} were derived from atomistic simulations. 
When averaged GB diffusivities derived from atomic GB models were used, the continuum simulations assumed \smash{$D^{\mathrm{GB}} = D^{\mathrm{GB}}_\parallel = D^{\mathrm{GB}}_\perp$}. 

The concentration field satisfies the steady-state balance equation,
\begin{align}
    \nabla\cdot\mathbf{q}(\mathbf{x}) = 0,
\end{align}
corresponding to the time-independent condition reached in the long-time limit. 
By imposing a concentration gradient and solving for the resulting concentration fluctuation field under periodic boundary conditions, the problem is treated using linear computational homogenization, yielding the macroscopic diffusivity tensor \smash{$\mathbf{D}^{\mathrm{macro}}$} for the RVE on the continuum scale~\cite{Scholz_Ou_Grabowski_Fritzen_2025}. 
Since the constructed polycrystalline model is nearly isotropic, only the average eigenvalue of the diffusivity tensor is reported and denoted as $D^{\mathrm{macro}}$. 
This average deviates from the minimum and maximum eigenvalues by at most 3\%, and typically by less than 1\%.
Comparisons with the analytical Maxwell--Garnett estimate~\cite{Garnett1904} for isotropic GB setups show good agreement. 
We emphasize that our approach extends the prediction of \smash{$D^{\mathrm{macro}}$} to polycrystalline solid electrolytes with transversely isotropic GBs. 
Schematic figures were created using Microsoft PowerPoint. 

In the resolved approach, the FFT--based homogenization solver \textsc{FANS}~\cite{Leuschner2018a} (see Code availability) was operated on a voxel-based image of the polycrystalline model.
For production runs, a $512\times512\times512$ voxel grid, corresponding to $\sim$134 million degrees of freedom, was used. 
In the collapsed approach, a conforming mesh was employed, with $\sim$0.44 million degrees of freedom for the 27-grain tessellation and $\sim$0.23 million degrees of freedom for the two-grain tessellation. 
The grain size of the polycrystalline model was defined as the diameter of a sphere with a volume equal to the average grain volume in the model. 
Grain-size variation in the polycrystalline model was achieved by rescaling the simulation cell size while keeping the GB width unchanged.

\bigskip

During the preparation of this manuscript, the authors used ChatGPT (OpenAI GPT-5 model) to improve clarity and readability. 
All content generated with its assistance was reviewed, edited, and verified by the authors, who take full responsibility for the final content of the published article.

\section*{Data availability}

The data generated in this study have been deposited in the DaRUS repository~\cite{Ou_2026_data}. 
The atomic dataset includes DFT and MD input files, training data, and fitted MTPs.
The continuum dataset contains the employed geometries, full-field solutions, and effective responses.
%
%
Source data are provided with this paper.

\section*{Code availability}

Scripts for performing quality-level-based local active learning, generating bulk, GB, and polycrystal structures, and analyzing results have been deposited in the DaRUS repository~\cite{Ou_2026_data}. 
An open-source and easy-to-use demonstrator for generating voxel-based geometries, running simulations with \textsc{FANS} and \textsc{MSUtils}, and visualizing results is provided on GitHub (\url{https://github.com/DataAnalyticsEngineering/PolycrystalDiffusion}). 
Additional codes are available from the corresponding authors upon reasonable request.

\onecolumngrid


\vspace{1cm}

\twocolumngrid

\makeatletter
\renewcommand\bibsection{%
  \section*{References}%
}
\makeatother

\bibliography{ref}

\onecolumngrid


\vspace{1cm}

\twocolumngrid

\begin{acknowledgments}

Y.O. thanks K. Jun, G. Winter, and X. Du from MIT for helpful discussions.

\end{acknowledgments}

\section*{Funding}

This project is funded by the Deutsche Forschungsgemeinschaft (DFG, German Research Foundation)--EXC 2075--390740016 under Germany’s Excellence Strategy (Y.O., L.S., S.K., Y.I., F.F., and B.G.); IK 125/1-1--519607530 (Y.I.); DI 1419/18-1--429191048 (S.D.); FR 2702/10--517847245 (F.F.); SFB 1333/2 2022--358283783 (B.G.); INST 40/575-1--405998092 FUGG (JUSTUS 2 cluster). 
Funding by the European Research Council (ERC) under the European Union’s Horizon 2020 Research and Innovation Program (Grant Agreement No.~865855) and under the European Union’s Horizon Europe Research and Innovation Programme (Grant Agreement No.~101200433, project META-LEARN) is acknowledged. 
Simulations were performed on the national supercomputer HPE Apollo Hawk at the High Performance Computing Center Stuttgart (HLRS) under the grant number H-Embrittlement/44239. 
The authors acknowledge the support by the Stuttgart Center for Simulation Science (SimTech), the state of Baden-Württemberg through bwHPC, and the MIT Office of Research Computing and Data for providing high-performance computing resources. 
The authors gratefully acknowledge the scientific support and HPC resources provided by the Erlangen National High Performance Computing Center (NHR@FAU) of the Friedrich-Alexander-Universität Erlangen-Nürnberg (FAU) under the NHR project~a102cb.
NHR funding is provided by federal and Bavarian state authorities. NHR@FAU hardware is partially funded by the DFG grant No.~440719683. 

Funded by the European Union. Views and opinions expressed are, however, those of the author(s) only and do not necessarily reflect those of the European Union or the European Research Council Executive Agency. Neither the European Union nor the granting authority can be held responsible for them.

\section*{Author contributions}

B.G., F.F., and Y.I. initiated, and Y.O. led the project;
Y.O. developed the machine-learning potentials and performed the atomic simulations under the supervision of B.G., Y.I., R.G.-B., M.K., W.Z., F.F., and S.D.;
L.S. and S.K. performed the continuum simulations under the supervision of F.F., B.G., and S.D.;
Y.O., L.S., S.K., Y.I., and M.K. analyzed the data. 
Y.O. wrote the paper with input from all authors.

\section*{Competing interests}

The authors declare no competing interests.


\end{document}


\title{\texorpdfstring{Supplementary Information for \\[5pt]
Microstructural Insights into Fast Ion Transport in Solid Electrolytes via Multiscale Modeling}{Supplementary Information for Microstructural Insights into Fast Ion Transport in Solid Electrolytes via Multiscale Modeling}}

\author{Yongliang Ou}
\email{yongliang.ou@imw.uni-stuttgart.de}
\affiliation{Institute for Materials Science, University of Stuttgart, Stuttgart, Germany}
\affiliation{Department of Materials Science and Engineering, Massachusetts Institute of Technology, Cambridge, MA, USA}

\author{Lena Scholz}
\email{scholz@mib.uni-stuttgart.de}
\affiliation{Institute of Applied Mechanics, University of Stuttgart, Stuttgart, Germany}

\author{Sanath Keshav}
\affiliation{Institute of Applied Mechanics, University of Stuttgart, Stuttgart, Germany}

\author{Yuji Ikeda}
\affiliation{Institute for Materials Science, University of Stuttgart, Stuttgart, Germany}

\author{Marvin Kraft}
\affiliation{Institute of Inorganic and Analytical Chemistry, University of Münster, Münster, Germany}
\affiliation{Institute of Energy Materials and Devices (IMD), IMD-4:~Helmholtz-Institut Münster, Forschungszentrum Jülich, Münster, Germany}

\author{Sergiy Divinski}
\affiliation{Institute of Materials Physics, University of Münster, Münster, Germany}

\author{Rafael Gómez-Bombarelli}
\affiliation{Department of Materials Science and Engineering, Massachusetts Institute of Technology, Cambridge, MA, USA}

\author{Wolfgang G. Zeier}
\affiliation{Institute of Inorganic and Analytical Chemistry, University of Münster, Münster, Germany}
\affiliation{Institute of Energy Materials and Devices (IMD), IMD-4:~Helmholtz-Institut Münster, Forschungszentrum Jülich, Münster, Germany}

\author{Felix Fritzen}
\email{fritzen@mib.uni-stuttgart.de}
\affiliation{Institute of Applied Mechanics, University of Stuttgart, Stuttgart, Germany}

\author{Blazej Grabowski}
\email{blazej.grabowski@imw.uni-stuttgart.de}
\affiliation{Institute for Materials Science, University of Stuttgart, Stuttgart, Germany}

\date{\today}

\maketitle

{
\onecolumngrid

\begin{center}

\vspace{-0.3cm}

\begin{minipage}{0.6\textwidth}
  \hypersetup{linkcolor=black} 
  \tableofcontents
\end{minipage}

\vspace{1.4cm}

{\small \bfseries List of supplementary items}
\vspace{0.8cm}

\hspace*{0.82cm}
\begin{minipage}{\linewidth}

\begin{minipage}[t]{0.42\textwidth}

\begin{itemize}[itemsep=0.1cm]
    \tableentry{tab:transferbility}
    \tableentry{tab:gb_ratio}
    \tableentry{tab:history}
    \tableentry{tab:allratio}
    \vspace{0.15cm}
    \figentry{fig:al_simple}
    \figentry{fig:al_full}
    \figentry{fig:al_detail}
    \figentry{fig:workflow}
    \figentry{fig:accuracy}
    \figentry{fig:efficiency}
    \figentry{fig:train_error}
    \figentry{fig:vali_error}
    \figentry{fig:literature}
\end{itemize}
\end{minipage}
\hspace{1.5cm}
\begin{minipage}[t]{0.42\textwidth}
\begin{itemize}[itemsep=0.115cm]
    \figentry{fig:chemspace}
    \figentry{fig:femmodel}
    \figentry{fig:gbwidth}
    \figentry{fig:non-arrhenius}
    \figentry{fig:dis_arrhenius}
    \figentry{fig:ordervsdisorder}
    \figentry{fig:singlegb}
    \figentry{fig:colvspol}
    \figentry{fig:chemspace2}
    \figentry{fig:disorder_trend}
    \figentry{fig:grainsize}
    \figentry{fig:ratio}
    \figentry{fig:poriosity}
\end{itemize}
\end{minipage}
\end{minipage}
\end{center}
\clearpage
}

\section{Parameterization of moment tensor potentials}
\label{sec:detalmtp}

Supplementary Fig.~\ref{fig:al_simple} shows an overview of the parameterization scheme for moment tensor potentials, with extended information shown in Suppl.~Fig.~\ref{fig:al_full}. 
The concept of quality-level-based active learning was proposed and tested in a previous study~\cite{Ou2024Nov} for the same material system, i.e., argyrodites. 
Initially, ab initio molecular dynamics simulations are performed on the ideal bulk structure. 
Considering the high computational cost of ab initio molecular dynamics simulations, the simulation temperature is set relatively high (up to the melting point), and the simulation time is set relatively short (a few picoseconds). 
The high-temperature molecular dynamics simulations enable a quick exploration of the phase space for a fixed chemical configuration by thermal vibrations, and the resulting trajectories provide the basis for the moment tensor potential training. 
The initial optimization of the moment tensor potential parameters is essential in calculating the extrapolation grade based on the D-optimality criterion~\cite{Podryabinkin2023Aug}. 
Since the ab initio molecular dynamics trajectories contain structurally similar configurations, only a small number of configurations from the ab initio molecular dynamics are chosen for the initial moment tensor potential training, and a pretraining step is introduced to include more configurations evaluated as extrapolation, i.e., with an extrapolation grade larger than one. 
Unlike standard active learning schemes, where a high level of moment tensor potential is directly used for active learning of complex structures, a low level of moment tensor potential is used for standard active learning, and the target material properties (for example, formation energy or diffusion coefficient) are evaluated at the end of active learning with the obtained low-level moment tensor potentials. 
If the accuracy of the target material properties is insufficient, a higher level of untrained moment tensor potential is used, and the pretraining, the standard active learning, and the evaluation processes are repeated. 
During the step-by-step training of moment tensor potentials with an increasing level, the configurations selected by active learning and labeled by density-functional theory are accumulated in the training set. 
The final moment tensor potential is obtained when the accuracy of the target material property is satisfied. 

In our previous paper~\cite{Ou2024Nov} focusing on grain boundaries in Li$_6$PS$_5$Cl, only three small grain boundaries were considered, and the full targeted grain boundary (GB) structure was used for active learning. This process is classified as global-AL. 
When the target structure is large, e.g., a large GB structure with more than \SI{30000}{} atoms, the density-functional theory calculation of the target structure selected by active learning becomes problematic. To overcome this problem, we have incorporated the quality-level-based concept with active learning of the local atomic environments (local-AL). The local-AL process proposed here has five main steps (Suppl.~Fig.~\ref{fig:al_detail}). 

(i)~Sampling. 
The moment tensor potential obtained after pretraining is used to run molecular dynamics simulations of the full target structure. 
For each snapshot resulting from molecular dynamics simulations, the extrapolation grade based on the local atomic environment is evaluated for each atom, and the maximum local grade of all atoms is focused on. 
Hyperparameters related to the threshold local grade, i.e., when to select the snapshot or stop the molecular dynamics simulations, need to be set. 
When the maximum local grade of a snapshot is larger than the set threshold, the snapshot is added to and accumulated in a sampling set. 

(ii)~Extraction. 
In this step, the number of atoms in the simulation cell, which is used for labeling (energy and forces calculation with density-functional theory) later on, is significantly reduced. 
First, the atom that has the maximum local grade is centered in the simulation cell. 
The local atomic environment of the centered atom is obtained by cutting out the atoms within a cubic box around the centered atom. 
The cubic box forms a new simulation cell with periodic boundary conditions. 
Next, atoms closely positioned due to the periodic boundary conditions are removed until the stoichiometry of atoms in the simulation cell is ensured. 

(iii)~Relaxation. 
This step aims to avoid unfavorable interactions of atoms at the boundary of the simulation box due to the periodic boundary conditions. 
To protect the local atomic environment that has been sampled, a protected region in the form of a cubic box is defined around the central atom, and the atoms within the protected region are fixed during the relaxation. 
The rest of the atoms in the simulation cell are relaxed to minimize the atomic forces. 
To reduce computational cost, density-functional theory calculations are performed with looser numerical convergence parameters. 
In this step, only the final relaxed structure enters the next step. 
Energies and forces obtained during structural relaxation do not enter the training set, maintaining the consistency of the data accuracy of the training set. 

(iv)~Labeling. 
The energies and forces of the relaxed structure are calculated by density-functional theory with tight convergence parameters. 
Because only single-point density-functional theory calculations are performed, the computational cost is kept low. 
The resulting energies and forces of all atoms in the simulation cell with the corresponding relaxed structures are added to and accumulated in the training set. 

(v)~Updating. 
The moment tensor potential, either initialized or obtained from the previous round of active learning, is retrained using the updated training set.
The resulting moment tensor potential is then used for the next round of local-AL, starting from Sampling. 

With this approach, active learning can be efficiently applied to large and structurally complex systems. 
Since ab initio molecular dynamics simulations are required only for the bulk unit cell, the overall computational cost is significantly reduced.
A schematic overview of the multiscale modeling framework incorporating local active learning is presented in Suppl.~Fig.~\ref{fig:workflow}.

{
\color{rev}

\section{Validation of moment tensor potentials}
\label{sec:valimtp}

The accuracy and efficiency of the quality-level-based local active learning scheme are demonstrated in Suppl.~Figs.~\ref{fig:accuracy} and~\ref{fig:efficiency} for a $\Sigma 3$ GB. 
Supplementary~Fig.~\ref{fig:train_error} shows the training energy and force errors obtained independently for various GB types ranging from $\Sigma 3$ up to $\Sigma 99$.
Although large grain boundaries such as $\Sigma 9$ and $\Sigma 11$ are structurally more complex than $\Sigma 3$ grain boundaries, only small increases are shown for training errors in energy (within \qty{2}{meV.atom^{-1}}) and force (within \qty{0.2}{eV.\angstrom^{-1}}). 
These results indicate that GB types exert only a limited effect on the training structures obtained via local active learning.

We also evaluated the accuracy of the moment tensor potentials for a $\Sigma 9$ GB and provide a comparison of the results to a $\Sigma 3$ GB in Suppl.~Fig.~\ref{fig:vali_error}. 
Moment tensor potentials reproduce the density-functional-theory-derived grain-boundary formation energies for both GB types. 
Additional density-functional-theory results also confirm that complex grain boundaries such as $\Sigma 9$ generally have higher formation energy than a $\Sigma 3$ GB. 
Next, we performed ab initio molecular dynamics simulations for the $\Sigma 3$ and $\Sigma 9$ grain-boundary structures at \qty{600}{\kelvin}, to generate configurations for validating forces predicted by the moment tensor potentials. 
Results show that errors for both grain boundaries are less than \qty{0.2}{eV.\angstrom^{-1}}, with the errors for $\Sigma 9$ only about \qty{0.03}{eV.\angstrom^{-1}} higher, confirming the high accuracy of the developed moment tensor potentials. 

Supplementary~Fig.~\ref{fig:literature} shows a comparison of results from moment tensor potentials and previous ab initio studies using the representative system Li$_6$PS$_5$Cl with 50\% anion disorder. 
We performed independent molecular dynamics simulations for six simulation models with random arrangement of Cl/S atoms. 
Different arrangements of Cl/S atoms were also considered for the RSC~Adv.~study~\cite{Lee_Lee_Park_Cho_Park_Sohn_2022}.
Diffusion barriers predicted from our work show much smaller variations due to different Cl/S arrangements (Suppl.~Fig.~\ref{fig:literature}\textbf{a}), likely due to the three orders of magnitude more atoms in the simulation cell and an order of magnitude longer simulation time compared to prior studies (Suppl.~Fig.~\ref{fig:literature}\textbf{c}).  
Nevertheless, the predicted diffusion barrier of \qty{213}{meV} from our study is within the range reported in previous ab initio studies (\qtyrange{202}{282}{meV})~\cite{Jiang_Chen_Rao_Chen_Zu_Singh_2022, Lee_Lee_Park_Cho_Park_Sohn_2022, Jeon_Cha_Jung_2024}. 
In all studies, diffusion coefficients at \qty{300}{\kelvin} were derived by extrapolating high-temperature diffusivities using Arrhenius relations (Suppl.~Fig.~\ref{fig:literature}\textbf{b}). 
Such extrapolation is sensitive to statistical errors of the Arrhenius fitting and availability of low-temperature data points.
The predicted diffusion coefficients are higher than previously reported values but remain within one order of magnitude. 
Discrepancies may arise from the limited length and time scales accessible in earlier ab initio studies.
We further validated the MTP-MD–predicted diffusivities by comparison with experimental values for anion-disordered Li$_6$PS$_5$Cl (Suppl.~Fig.~\ref{fig:chemspace}).
Our results show good agreement with quasielastic neutron scattering measurements.

The transferability of moment tensor potentials developed via local active learning was evaluated across the anion ordering chemical space in argyrodites (Suppl.~Table~\ref{tab:transferbility}).
Transferability is relatively good for Li$_6$PS$_5$Cl, whereas it is limited for Li$_6$PS$_5$Br and Li$_6$PS$_5$I. 

}

\section{Consistency of continuum simulation models}
\label{sec:consistency}

\subsection{Methodology}

We employed two representations for GBs in polycrystalline solid electrolytes: (i)~a volume-resolved GB model, in which the GB domain is explicitly meshed (Suppl.~Fig.~\ref{fig:femmodel}\textbf{a}), and (ii)~a thickness-collapsed model, in which two-dimensional submanifolds represent GBs (Suppl.~Fig.~\ref{fig:femmodel}\textbf{b}). 
The volume-resolved approach is feasible only when the GB width exceeds the numerical mesh size and when sufficient spatial resolution is available to capture the behavior within the GB domain. 
Hence, this model is limited to nanograin setups. 
In contrast, the collapsed GB model assumes a thin GB---typical for micrograins---so there is a clear separation of length scales for grain size and GB width. 
Due to the conceptually different assumptions concerning the GB representation, a straightforward quantitative comparison of their results is non-trivial. 
We therefore systematically studied the respective applicability regimes required to guarantee physically sound results. 
In particular, a transition grain size was determined at which one should switch from the volume-resolved to the thickness-collapsed model.

To isolate the intrinsic behavior of the two continuum models and eliminate any geometrical artifacts, we employed a highly idealized microstructure:~a two-grain Voronoi tessellation. 
This configuration is perfectly isotropic, features symmetric GB junctions, and keeps the computational cost low. 
The collapsed approach allows for a parametric variation of the GB width on a single mesh. 
In contrast, changing the GB width in a resolved model requires rebuilding the three-dimensional voxel image each time to properly identify GB voxels. 
Therefore, we generated resolved polycrystalline models for a fixed spatial resolution where the GB width is a multiple of the voxel size. 
Because the considered Voronoi tessellation was fixed, increasing the GB width reduces the effective grain size. 
Designing a corresponding collapsed model that matches the GB width and GB volume fraction of the resolved model is then straightforward. 
This was done by rescaling the GB width parameter together with the simulation cell length. 
Simulations were then carried out for both GB representations independently for a range of bulk and GB diffusivities, $D^{\mathrm{bulk}}$ and \smash{$D^{\mathrm{GB}}:=D^{\mathrm{GB}}_\parallel=D^{\mathrm{GB}}_\perp$}. 
By repeating the procedure at several mesh resolutions, we could (i)~compare the results of both models across different grain size regimes depending on the inputs $D^{\mathrm{GB}}/D^{\mathrm{bulk}}$, and (ii)~assess how the spatial resolution influences the accuracy of the resolved approach. 
From these systematic studies, we (i)~identified the transition grain size at which the modeling strategy should be switched (\ref{sec:conti-gs}), and (ii)~formulated a practical guideline for selecting a mesh resolution that ensures convergence and accurate results for the resolved approach (\ref{sec:conti-resol}).

\subsection{Transition grain size and agreement of the models}
\label{sec:conti-gs}

The true diffusion behavior of a polycrystalline material cannot be measured in detail, so the following considerations are based on two assumptions: (i)~the resolved approach yields accurate results in the nanograin regime, and (ii)~the collapsed model reliably reproduces diffusion in the micrograin regime. 
When both models are applied across the entire range of grain sizes, we therefore expect a transition point in between the nanograin and micrograin limits where the agreement is significantly improved compared to the two extremes. 
This is supported by the relatively small deviation in macroscopic diffusivity for medium-sized grains (Suppl.~Fig.~\ref{fig:femmodel}\textbf{c}).
Consequently, we define the transition grain size as the grain size at which the two models are most consistent, regardless of the GB-to-bulk diffusion contrast $D^{\mathrm{GB}}/D^{\mathrm{bulk}}$, i.e., we minimize the sum of squared relative deviations. 
The transition grain size (indicated by the dashed lines in Suppl.~Fig.~\ref{fig:femmodel}\textbf{c}) depends strongly on the spatial resolution of the resolved model, which will be discussed in \ref{sec:conti-resol}. 
Upon mesh refinement, the transition size converges to about 100~nm. 
This value was therefore adopted as a reference for all continuum-model results presented in this study. 
For the computations that cover the entire grain size range (e.g.,~Suppl.~Fig.~\ref{fig:grainsize}), both the resolved and the collapsed representations were evaluated slightly beyond the identified transition grain size to demonstrate the consistency of the two approaches.

The obtained transition grain size of 100~nm is specific to the considered GB width of 2.5~nm that was assumed for the given geometry. 
Nevertheless, the same procedure can be applied to any GB width or microstructure. 
Converting the transition grain size to a GB volume fraction yields approximately 6.3\%, which serves as a GB width-independent indicator of the regime change. 

Even in the vicinity of the transition grain size, the two approaches do not achieve perfect agreement (Suppl.~Fig.~\ref{fig:femmodel}\textbf{c}). 
This residual mismatch stems from inherent differences between the two GB representations: 
(i)~the explicit representation of GB junction domains in the volume-resolved approach versus their implicit treatment in the collapsed approach, 
(ii)~the volume corrections required for the collapsed approach, and 
(iii)~the underlying microstructures considered for the two approaches are not identical but only match key characteristics. 
Nevertheless, the deviation is modest---typically below 5\%---at the transition grain size. 

Large deviations occur primarily for blocking GBs, i.e., when the GB diffusivity is lower than the bulk diffusivity ($D^{\mathrm{GB}} < D^{\mathrm{bulk}}$). 
In this regime, the less-conductive GBs induce pronounced concentration jumps across the GB. 
These jumps and their effects are especially difficult to capture if not resolved well, as in the following two cases: 
(i)~in volume-resolved simulations of micrograins, a GB may be represented by only a few voxel layers, leading to an under-resolved GB domain and noticeable staircasing effects; and
(ii)~in the collapsed approach for nanograins, the zero-width assumption neglects a substantial GB volume. 
The impact of resolution in case (i) on concentration and flux fields is shown in Suppl.~Fig.~\ref{fig:femmodel}\textbf{d} for the smallest GB width relative to the grain size. 
Notably, the volume fraction varies substantially with resolution, even when the GB width remains fixed. 
This leads to differences in the flux magnitude in the presented results obtained with the volume-resolved approach. 

When GB diffusion dominates in the polycrystalline model ($D^{\mathrm{GB}} > D^{\mathrm{bulk}}$), results from the volume-resolved and thickness-collapsed approaches converge for micrograins. 
In this diffusion regime, GBs do not cause concentration discontinuities but serve as additional pathways.
Because the grains are large, the contribution of these pathways to the overall transport is modest, so the agreement between the two models persists even when the GB width is resolved by only a few voxel layers. 
Therefore, the volume-resolved approach can be employed reliably for grain sizes beyond the estimated transition grain size (100~nm).

\subsection{Resolution requirements for volume-resolved grain boundaries}
\label{sec:conti-resol}

When a resolved microstructure is generated for a prescribed GB width for different resolutions, the resulting GB volume fraction varies slightly because the voxel-based discretization only approximates the GB domain. 
This prevents a direct comparison of the macroscopic diffusivities across different resolutions, since the GB volume fraction itself influences the resulting macroscopic diffusivities. 
Therefore, the convergence behavior of the volume-resolved approach cannot be evaluated unambiguously. 
Supplementary~Fig.~\ref{fig:femmodel}\textbf{c} shows that the mesh resolution has a noticeable impact on the accuracy of the results, considering the corresponding results from the collapsed approach as a resolution-independent reference. 

In the polycrystalline model, the relative GB width decreases with increasing grain size. 
Consequently, for a given resolution, the number of voxel layers representing a GB varies with grain size. 
Supplementary~Fig.~\ref{fig:femmodel}\textbf{c} shows that only a few layers were used to resolve a GB for micrograins, and more layers were used for nanograins. 
The resolution effect is most pronounced for blocking GBs ($D^{\mathrm{GB}}<D^{\mathrm{bulk}}$) combined with large grains, i.e., when the resolved model under-resolves the GB. 
The convergence of the relative error curves with increasing resolution implies a practical rule:~The GB width should be at least ten times the voxel length. 
For a given geometry, the GB width in the unit cube is first adjusted to satisfy the prescribed ratio between GB width and grain size.
The resolution is then selected such that the voxel size is less than one-tenth of the GB width. 
In contrast, as discussed in \ref{sec:conti-gs}, results with an enhancing GB setup ($D^{\mathrm{GB}}>D^{\mathrm{bulk}}$) exhibit almost no sensitivity to mesh resolution. 
In those cases, a coarser discretization is sufficient:~A GB width of more than five times the voxel length yields mesh-independent results. 
In the present study, all volume-resolved simulations were conducted on a $512\times512\times512$ voxel grid, which exceeds the resolution required to resolve GBs in polycrystalline models with nanograins (grain size less than 100~nm) while avoiding prohibitive computational cost.

{\color{rev}

\section{Sensitivity to grain boundary width}
\label{sec:sen_width}

Sensitivity to the GB width can be straightforwardly quantified within our multiscale modeling framework.
At the atomistic scale, GB diffusivities derived from polycrystalline models depend on the assumed GB width.
For a two-grain polycrystalline model with cubic geometry, the chosen GB widths and the corresponding calculated GB volume fractions are summarized in Suppl.~Table~\ref{tab:gb_ratio}.
Supplementary~Fig.~\ref{fig:gbwidth} shows the predictions obtained using different GB widths across atomistic and continuum scales.
Reducing the GB width from \qty{2.5}{nm} to \qty{1.5}{nm} yields similar Li-ion diffusion barriers, consistent with experimental measurements. 
In contrast, a further reduction to \qty{1}{nm} leads to an overestimation of the diffusion barriers relative to experiment.
For polycrystals with small grains, diffusivities obtained from all-atom MD simulations agree well with those predicted by our multiscale framework using a GB width of \qty{2.5}{nm}.
Smaller widths of \qty{1}{nm} and \qty{1.5}{nm} introduce inconsistencies between atomistic and continuum-scale simulations.

\section{Non-Arrhenius transport behavior}
\label{sec:nonarr}

Non-Arrhenius behavior is observed for anion-ordered Li$_6$PS$_5$I and Li$_6$PS$_5$Br, as shown in Suppl.~Fig.~\ref{fig:non-arrhenius}. 
At higher temperatures, the diffusivity ratios between GB and bulk are relatively small, and the total diffusivity is dominated by the bulk diffusivity due to the large volume fraction of bulk in the polycrystal model. 
At lower temperatures, the diffusivity ratios between GB and bulk become larger, and the effect of GBs becomes more critical.
Therefore, a grain-boundary-dominated regime is found at low temperatures. 
Such a GB-induced non-Arrhenius behavior was reported in a previous study~\cite{Heo2021Dec}.

On the other hand, the non-Arrhenius behavior cannot be clearly observed for anion-disordered argyrodites in the investigated temperature range, as demonstrated for Li$_6$PS$_5$Cl with 50\% anion disorder in Suppl.~Fig.~\ref{fig:dis_arrhenius}.  
The bulk and GB diffusivities are similar, even in the low-temperature range, so the polycrystal diffusion behavior closely follows the bulk Arrhenius fit, without showing non-Arrhenius behavior in the investigated temperature range.

Supplementary~Fig.~\ref{fig:ordervsdisorder} compares predicted temperature-dependent diffusivities of different argyrodite polycrystals. 
Note that the non-Arrhenius behavior, including the transition temperature, is also dependent on the grain size of the polycrystal model. 
Our results are consistent with experimental observations, which report non-Arrhenius diffusion behavior in Li$_6$PS$_5$I but not in Li$_6$PS$_5$Br~\cite{Brinek2020Jun}. 
In experiments, Li$_6$PS$_5$I is predominantly anion-ordered, whereas Li$_6$PS$_5$Br is generally anion-disordered.
Further application of our multiscale modeling framework enables a priori screening of material systems that show anomalous diffusion behavior. 
    
}

\section{Additional results and discussion}

An overview of investigations on GBs in solid electrolytes is presented in Suppl.~Table~\ref{tab:history}.
The present study extends previous investigations to a systematic class of sulfides and a broader set of GBs using accurate machine-learning potentials.
Comparisons of Li-ion diffusivities for various single-type GBs and between columnar and polycrystalline grains are shown in Suppl.~Figs.~\ref{fig:singlegb} and \ref{fig:colvspol}, respectively.
GB effects across the chemical space of argyrodites, including anion species and ordering, are shown in Suppl.~Figs.~\ref{fig:chemspace2} and \ref{fig:disorder_trend}.
The predicted Li-ion diffusivities from this study are summarized in Suppl.~Table~\ref{tab:allratio}, and the derived hopping frequencies are consistent with experimental values~\cite{shotwell_tetrahedral_2025}.
Using the multiscale modeling framework, grain-size-dependent diffusivities of polycrystalline argyrodites (Suppl.~Fig.~\ref{fig:grainsize}) and a precomputed map linking atomic-scale mechanisms to continuum-scale behavior in general polycrystalline solid electrolytes (Suppl.~Fig.~\ref{fig:ratio}) are obtained. 
Moreover, the impact of porosity on macroscopic conductivity can be incorporated on top of these results, as demonstrated for anion-disordered Li$_6$PS$_5$Cl (Suppl.~Fig.~\ref{fig:poriosity}).

\onecolumngrid

\clearpage
\begin{supfigure*}
    \capstart
    \centering
    \includegraphics{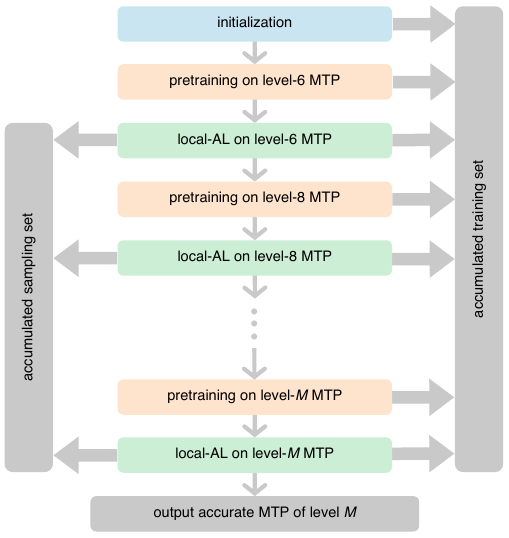}
    \caption[Quality-level-based MTP training scheme]{
    \textbf{
    Overview of the parameterization scheme for moment tensor potentials (MTP).}
    The quality-level-based concept is emphasized with the changing MTP level.
    }
    \label{fig:al_simple}
\end{supfigure*}

\newpage
\begin{supfigure*}
    \capstart
    \centering
    \includegraphics{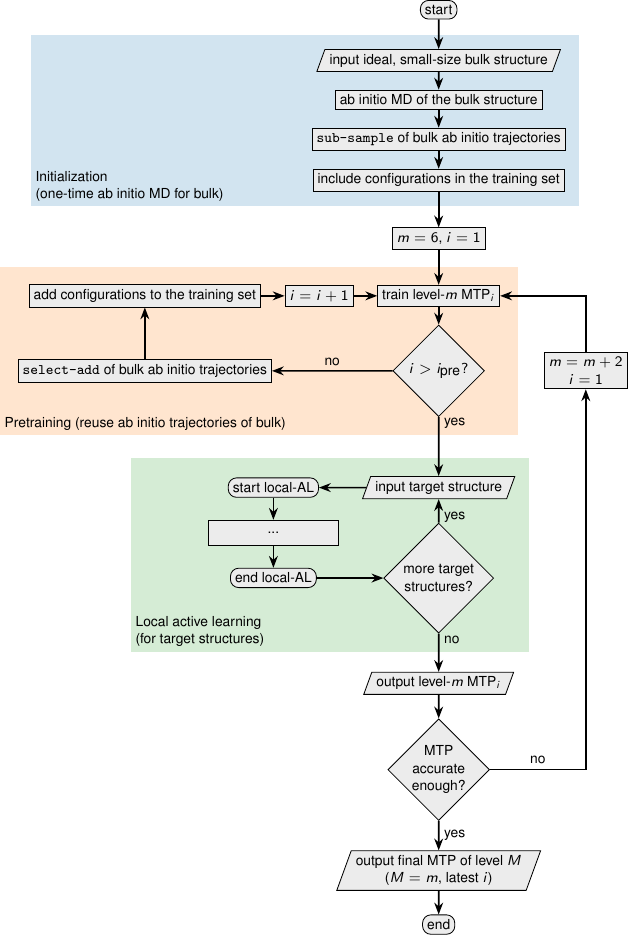}
    \caption[Complete iterative MTP training workflow]{
    \textbf{Complete workflow of the parameterization scheme for moment tensor potentials (MTPs).}
    MD stands for molecular dynamics simulations, while AL stands for active learning. 
    Details of the local active learning steps, indicated by “$\cdots$” in the present figure, are provided in Suppl.~Fig.~\ref{fig:al_detail}.
    }
    \label{fig:al_full}
\end{supfigure*}

\newpage
\begin{supfigure*}
    \capstart
    \centering
    \includegraphics{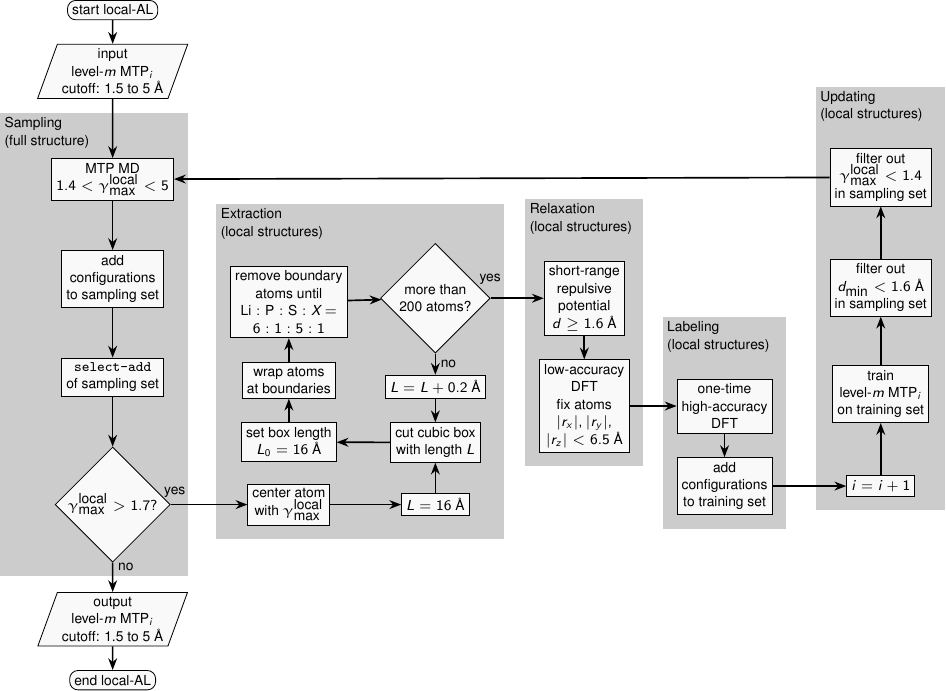}
    \caption[Steps and hyperparameters of local AL]{
    \textbf{Details of active learning (AL) for local structures.}
    Parameters used for Li$_6$PS$_5X$ with $X\in \{\mathrm{Cl}, \mathrm{Br}, \mathrm{I} \}$ are shown. MTP-MD and DFT refer to molecular dynamics simulations with moment tensor potentials and density-functional theory, respectively. The extrapolation grade and atomic distance are denoted by $\gamma$ and $d$, respectively. 
    }
    \label{fig:al_detail}
\end{supfigure*}

\newpage
\begin{supfigure*}[!htbp]
    \capstart
    \centering
    \includegraphics{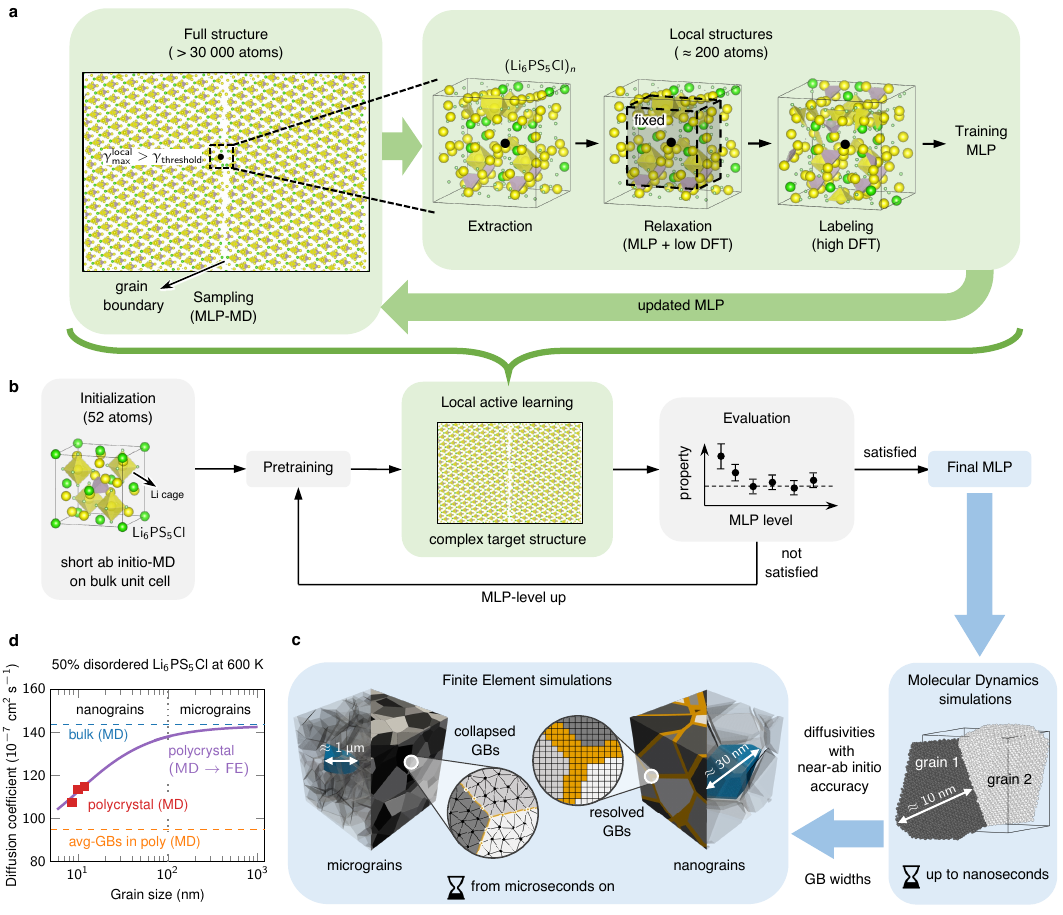}
    \caption[Overview of multiscale modeling framework]{
    \textbf{Schematic illustration of the multiscale modeling framework for resolving Li-ion transport in polycrystalline solid electrolytes.} 
    \textbf{a}, Closed-loop local active learning. 
    Local structures are extracted from the full structure for efficient labeling with density-functional theory (DFT). 
    Sampling is based on the local grade (\smash{$\gamma^{\mathrm{local}}$}) from the machine-learning potentials (MLPs). 
    The extracted local structure retains its stoichiometry. 
    Active learning finishes when no more configurations are sampled from MD simulations. 
    \textbf{b}, Quality-level-based parameterization. 
    Accurate and efficient MLP fitting is performed for complex structures, illustrated here with ordered Li$_6$PS$_{5}$Cl as an example. 
    The $\Sigma99[110]$ grain boundary (GB) embedded in a bicrystalline model is an example of the target structure. 
    After initialization, a low level of MLP is used for pretraining and active learning. 
    Then, the quality of the obtained MLP is evaluated based on the material properties of interest. 
    If the MLP quality is not satisfied, a higher level of MLP is used for the next stage of active learning. 
    \textbf{c}, Connecting atomic-scale to continuum-scale simulations. 
    Diffusivities with near-ab initio accuracy from MD simulations are inputs for FE simulations. 
    Different approaches are employed on the continuum scale to simulate GBs in polycrystals with nanograins and micrograins. 
    \textbf{d}, Example application of the multiscale modeling framework (MD~$\rightarrow$~FE) for predicting diffusivities of polycrystalline anion-disordered Li$_6$PS$_5$Cl at 600~K. 
    Results from MD simulations of polycrystalline models are compared with FE simulations to demonstrate their consistency. 
    }
    \label{fig:workflow}
\end{supfigure*}

\newpage
\begin{supfigure*}
    \capstart
    \centering
    \includegraphics{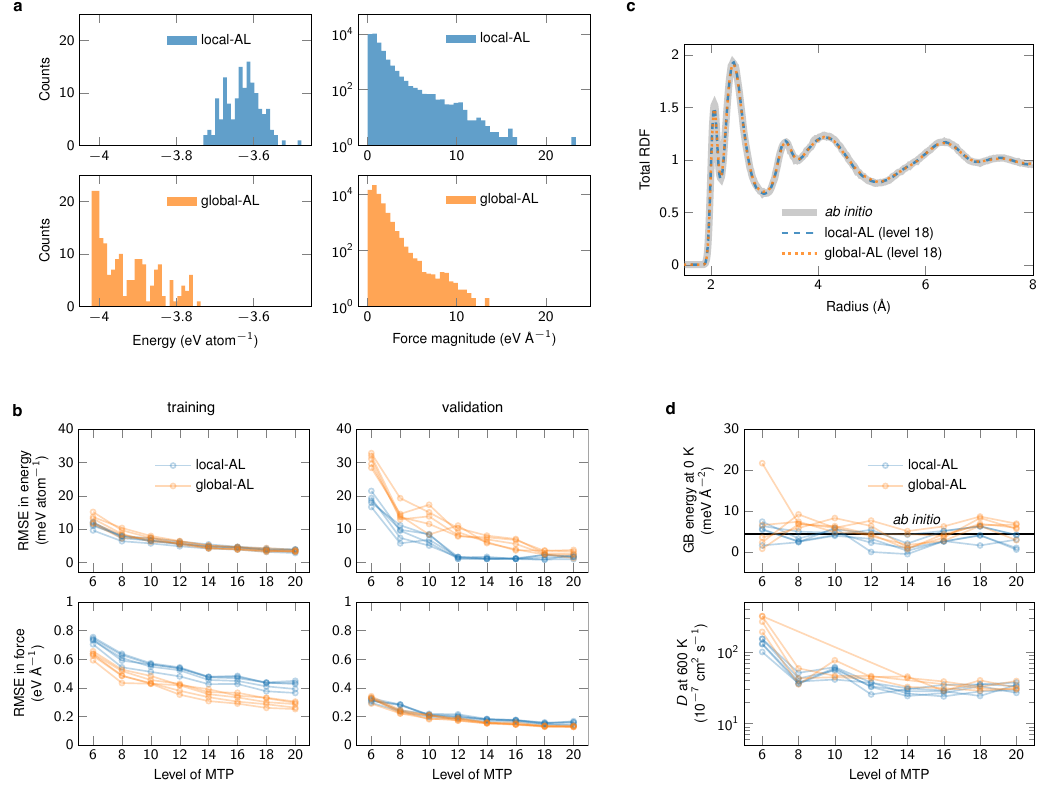}
    \caption[Accuracy of local active learning]{
    \textbf{Accuracy of the quality-level-based local active learning scheme.} 
    \textbf{a}, Energy and force distributions of configurations sampled using local- or global-active learning (AL) schemes up to level 20. 
    The term global-AL refers to cases where uncertainty is quantified over the entire simulation model, whereas local-AL refers to cases where uncertainty is quantified individually for each atom in the model. 
    The configurations used to train moment tensor potentials (MTPs) in the local-AL scheme are extracted local structures and thus exhibit higher potential energies. 
    \textbf{b}, Root mean square errors (RMSEs) of the trained MTPs from level 6 to 20 for the training and the validation sets. 
    For MTPs resulting from the local-AL scheme, smaller RMSEs are observed in potential energy for the validation set, while larger RMSEs are seen in atomic force for both the training and validation sets. 
    The validation set is obtained by MD simulations of grain boundary (GB) structures at 600 K. 
    \textbf{c}, Time-averaged radial distribution functions (RDFs) of the configurations obtained by running ab initio-molecular dynamics simulations and molecular dynamics simulations with level-18 MTPs for the $\Sigma3$ GB in anion-ordered Li$_6$PS$_5$Cl. 
    The MTPs were trained using the quality-level-based scheme with local- or global-AL. 
    Good agreement is observed between ab initio- and MTP-derived RDFs. 
    \textbf{d}, Comparison of the predicted GB formation energy and diffusivity for the $\Sigma3$ GB, using MTPs trained with local- or global-AL. 
    The GB energy from ab initio density-functional theory calculations is compared. 
    MTPs trained to higher levels (e.g., level 18) yield consistent results for both AL schemes. 
    To obtain statistics, five MTPs were independently trained with the same AL scheme and subsequently employed for RMSE and property analysis. 
    }
    \label{fig:accuracy}
\end{supfigure*}

\newpage
\begin{supfigure*}
    \capstart
    \centering
    \includegraphics{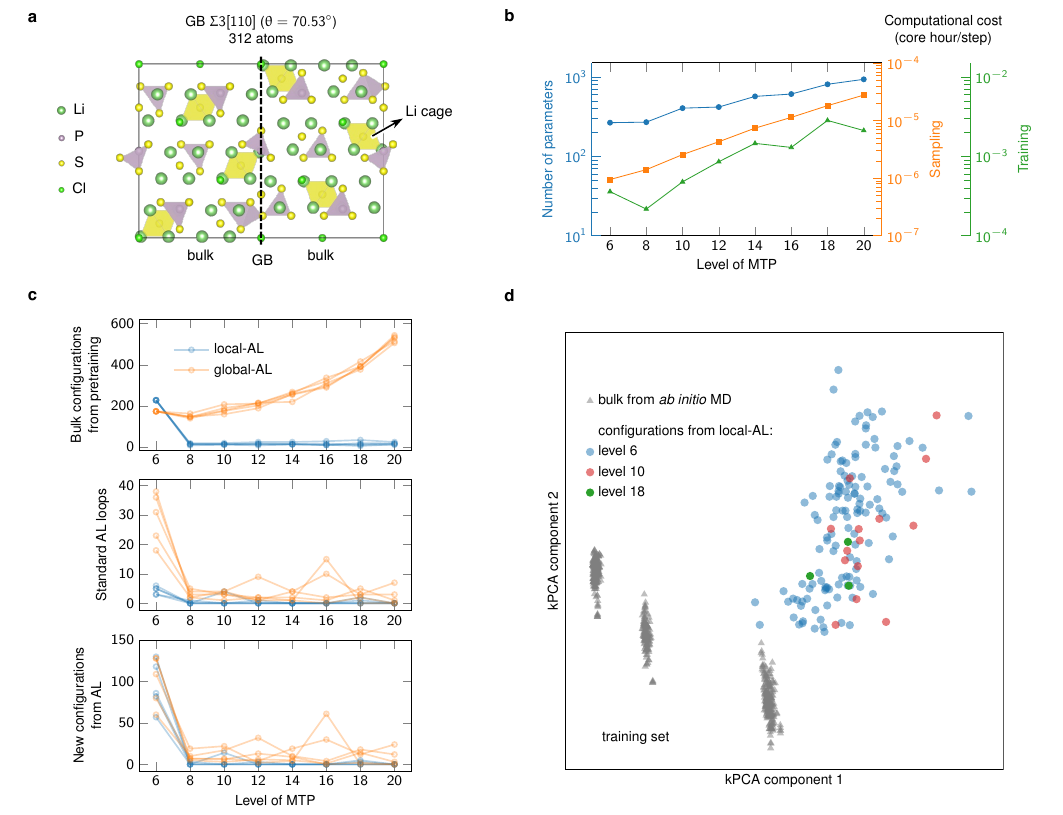}
    \caption[Efficiency of local active learning]{
    \textbf{Efficiency of the quality-level-based local active learning scheme.} 
    \textbf{a}, Atomistic structure of a bicrystalline model consisting of 312 atoms, incorporating single-type $\Sigma3$ grain boundaries (GBs). 
    The anion-ordered bulk structures of Li$_6$PS$_5$Cl were rotated by $\theta=\SI{70.53}{\degree}$ in the [110] direction to construct the GB. 
    The model is set as the target structure for comparison of local- and global-active learning (AL). 
    \textbf{b}, Number of fitting parameters in moment tensor potentials (MTPs) from level 6 to 20, along with the computational costs associated with MTP-based molecular dynamics simulations of the target structure (sampling) and MTP training. 
    All of them increase exponentially with the increasing level of MTP.  
    \textbf{c}, Comparison of the performance between local-AL and global-AL schemes for the targeted bicrystalline model. 
    At the end of each level of AL, the number of bulk configurations added to the training set during pretraining, the number of standard-AL loops performed, and the number of new configurations sampled and labeled during AL are presented. 
    Fewer configurations are needed to train the MTPs using the local-AL scheme. 
    Each AL scheme was performed five times for statistics. 
    \textbf{d}, Kernel principal component analysis (kPCA) of the training set after local-AL up to level 20. 
    More diverse configurations were sampled at lower levels of AL compared to higher levels. 
    Structural representations were constructed using smooth-overlap-of-atomic-positions (SOAP)~\cite{De_Bartók_Csányi_Ceriotti_2016} descriptors. 
    }
    \label{fig:efficiency}
\end{supfigure*}

\newpage
\begin{supfigure*}[!ht]
    \capstart
    \centering
    \includegraphics{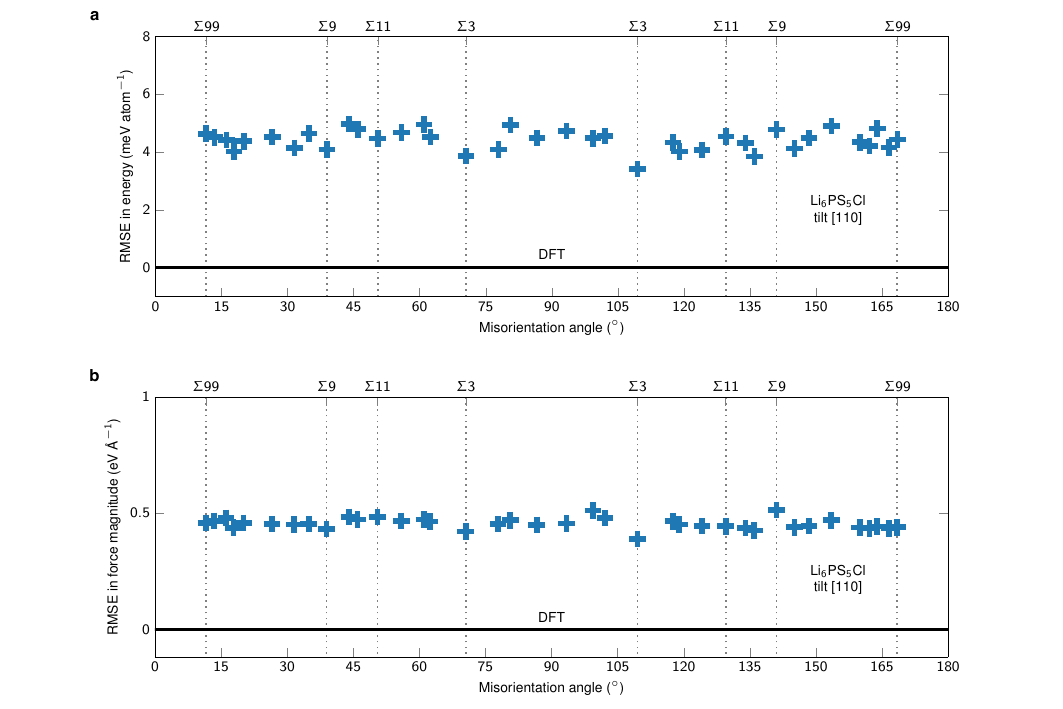}
    \caption[Training errors for single-type GBs]{
    \textbf{Energy and force errors for training structures obtained through local active learning.}
    \textbf{a}, Root-mean-square errors (RMSEs) in machine-learning-potential-predicted energies are around \qty{4}{meV.atom^{-1}}.
    \textbf{b}, RMSEs in force are around \qty{0.5}{eV.\angstrom^{-1}}.
    Errors in energies and forces remain consistent across different grain-boundary types.
    Results are based on single-type tilt grain boundaries in anion-ordered Li$_6$PS$_5$Cl with a [110] rotation axis. 
    }
    \label{fig:train_error}
\end{supfigure*}

\clearpage
\begin{supfigure*}[!ht]
    \capstart
    \centering
    \includegraphics{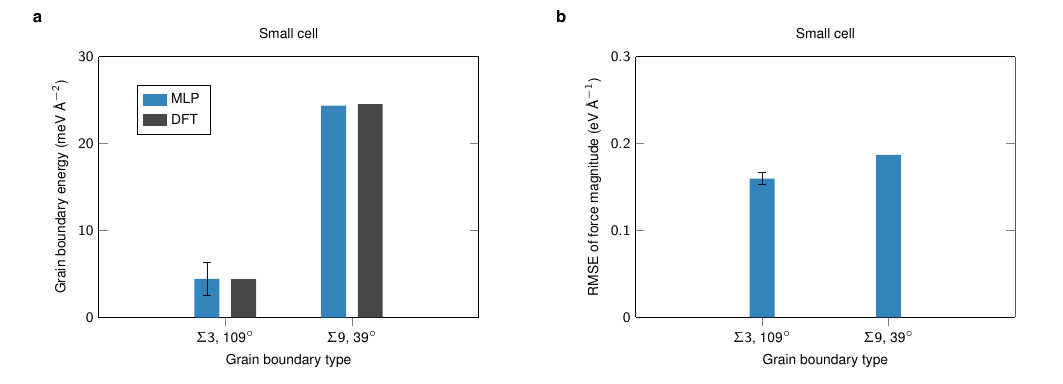}
    \caption[Validation errors for single-type GBs]{
    \textbf{Validation of machine-learning potentials (MLPs) for different types of grain boundaries.}
    Results for representative $\Sigma 3$ and $\Sigma 9$ grain boundaries in Li$_6$PS$_5$Cl, with misorientation angles of about \qty{109}{\degree} and \qty{39}{\degree}, respectively, are compared.
    \textbf{a}, Grain-boundary energetics at \qty{0}{\kelvin}. 
    MLPs reproduce grain-boundary formation energies obtained from density-functional theory (DFT). 
    $\Sigma 9$ grain boundaries exhibit formation energies about \qty{20}{meV.\angstrom^{-2}} higher than $\Sigma 3$.
    Error bars represent standard deviation of results from five independently trained potentials.
    \textbf{b}, Force prediction errors at \qty{600}{\kelvin}.
    Ab initio molecular dynamics simulations of grain-boundary structures at \qty{600}{\kelvin} were used to generate validation configurations. 
    Force errors for both $\Sigma 3$ and $\Sigma 9$ grain boundaries remain below \qty{0.2}{eV.\angstrom^{-1}}.
    Due to the high computational cost of DFT, grain-boundary structures were modeled in small simulation cells.
    Error bars represent the standard deviation across five independently trained MLPs obtained via local active learning.
    }
    \label{fig:vali_error}
\end{supfigure*}

\newpage
\begin{supfigure*}[!ht]
    \capstart
    \centering
    \includegraphics{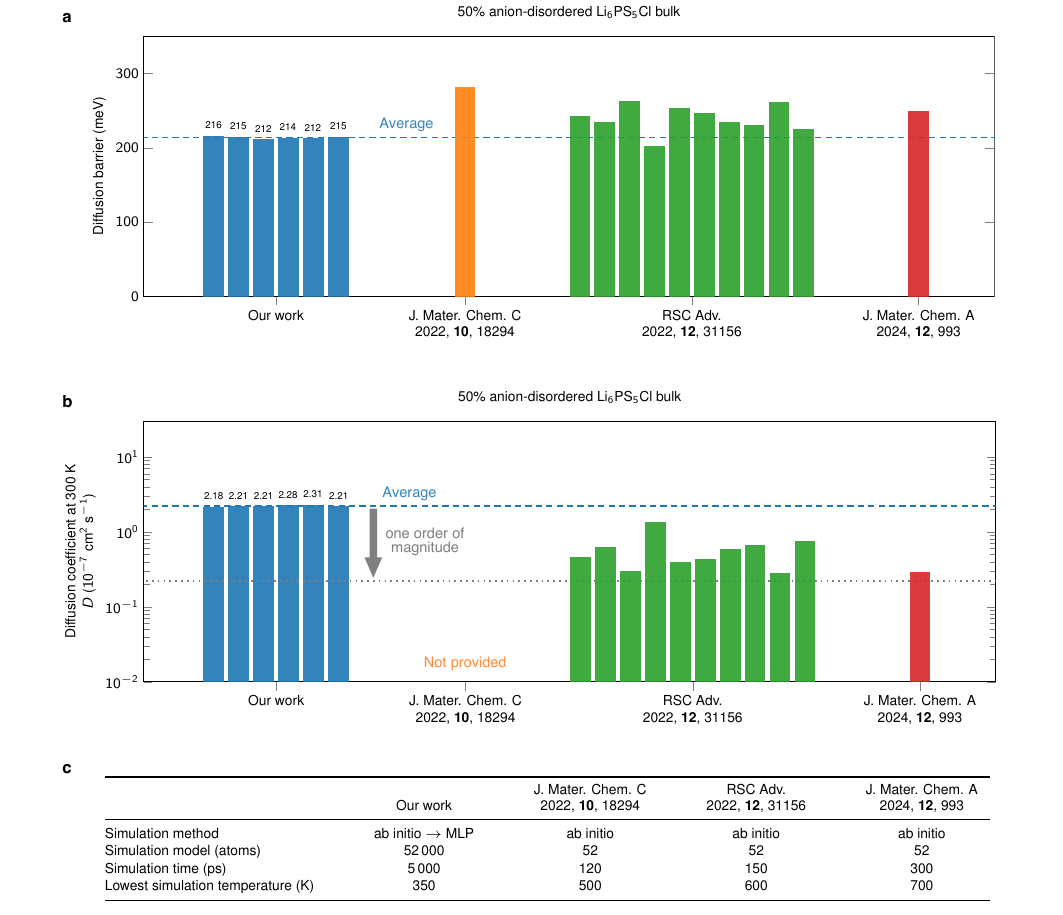}
    \caption[Comparison of MTPs to ab initio results]{
    \textbf{Comparison of MLP--MD-predicted diffusion properties with ab initio results.}
    In all studies, molecular dynamics simulations are performed and diffusion properties are extracted from the mean-squared displacement of Li ions. 
    Bulk Li$_6$PS$_5$Cl with 50\% anion disorder is considered.
    \textbf{a}, Comparison of diffusion barriers.
    \textbf{b}, Comparison of diffusion coefficients extrapolated to \qty{300}{\kelvin}.
    Dashed lines indicate averaged values from our work and facilitate comparison with previous studies (J.~Mater.~Chem.~C~\cite{Jiang_Chen_Rao_Chen_Zu_Singh_2022}, RSC~Adv.~\cite{Lee_Lee_Park_Cho_Park_Sohn_2022}, J.~Mater.~Chem.~A~\cite{Jeon_Cha_Jung_2024}).
    A dotted line indicates a value one order of magnitude lower than our predicted diffusivities for reference.
    \textbf{c}, Comparison of simulation length and time scales, together with the lowest temperatures sampled.
    Larger simulation models are essential to capture realistic anion disorder, whereas longer simulation times are required to obtain statistically converged diffusivities based on the Einstein relation.
    Additionally, access to lower temperatures is critical for reliable Arrhenius extrapolation to obtain room-temperature diffusivities.
    }
    \label{fig:literature}
\end{supfigure*}

\newpage
\begin{supfigure*}
    \capstart
    \centering
    \includegraphics{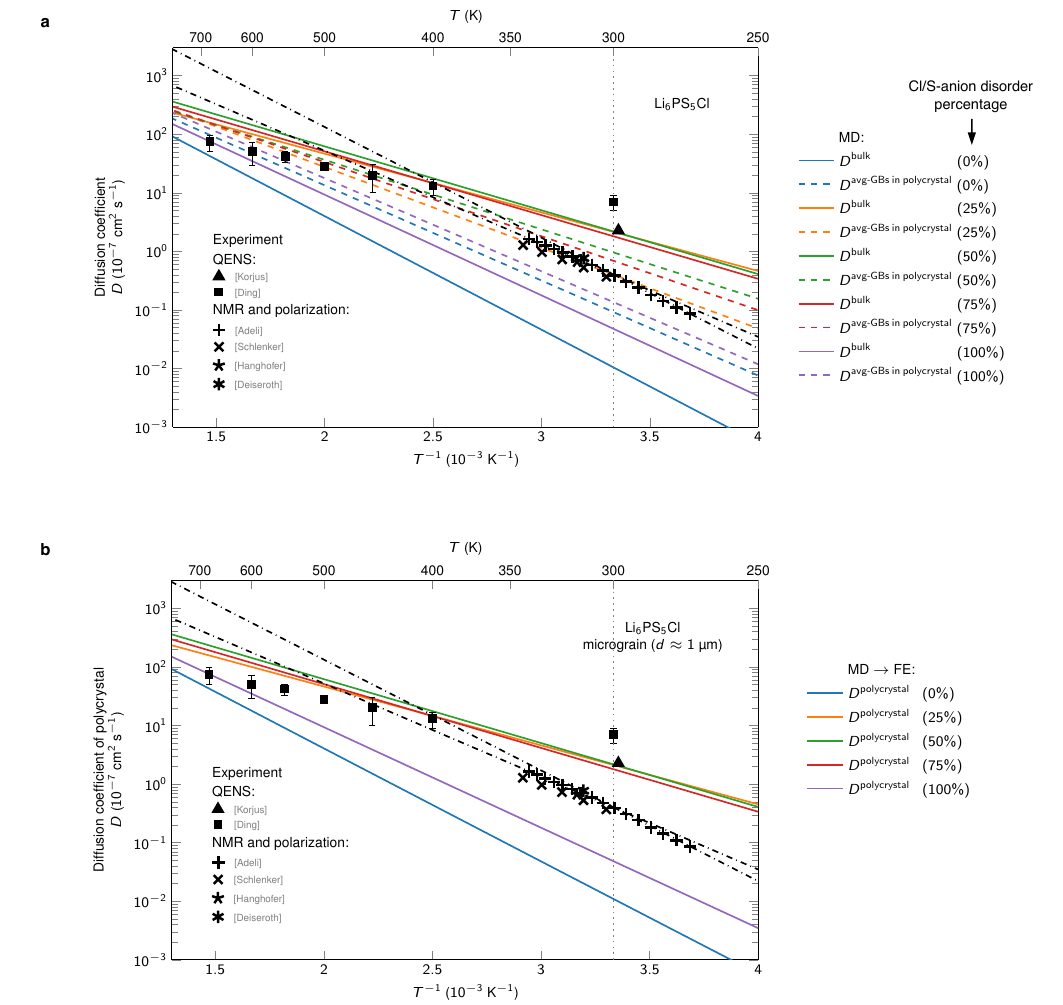}
    \caption[Li$_6$PS$_5$Cl with varying disorder ratios]{
    \textbf{Grain-boundary (GB) effects in Li$_6$PS$_5$Cl at varying anion-disorder ratios.}
    \textbf{a}, Arrhenius fits of predicted diffusion coefficients of bulk and GB structures in Li$_6$PS$_5$Cl with varying percentages of Cl/S-anion disorder. 
    The GB values (avg-GBs in polycrystal) were extracted from molecular dynamics simulations of polycrystalline models with random anion disorder. 
    Experimentally measured diffusion coefficients and their uncertainties from~\citet{Adeli2019Jun,Schlenker2020Oct,Hanghofer2019,Deiseroth2011Aug} (nuclear magnetic resonance and polarization measurements), and~\citet{Korjus_Mitra_Berrod_Vanpeene_Appel_Broche_Lyonnard_Villevieille_2025,Ding2025Jan} (quasielastic neutron scattering measurements), with samples exhibiting $\sim$60\% anion disorder, are compared. 
    \textbf{b}, Predicted diffusion coefficients of polycrystalline Li$_6$PS$_5$Cl with an average grain size of \SI{1}{\micro\meter}. 
    Experimental diffusion coefficients fall within the predicted diffusivity domains introduced by chemical and microstructural variations. 
    }
    \label{fig:chemspace}
\end{supfigure*}

\clearpage
\begin{suptable*}[!ht]
    \capstart
    \caption[Transferability of MTPs from local AL]{
    \textbf{Transferability of moment tensor potentials developed via local active learning.} 
    Moment tensor potentials trained on polycrystalline models with 50\% anion disorder are validated against bulk configurations with different degrees of anion disorder.
    The validation dataset is generated from molecular dynamics simulations at \qty{600}{\kelvin}.
    For Li$_6$PS$_5$Cl, the potentials show good agreement with reference energies and forces across all considered disorder configurations.
    In contrast, the potentials for Li$_6$PS$_5$Br and Li$_6$PS$_5$I exhibit significantly larger errors in both energies and forces.
    }
    \label{tab:transferbility}
    \vspace{0.1cm}
    \sffamily
        \sansmath
        \selectfont
\begin{ruledtabular}
\begin{tabular}{lccccc}
 System & Li$_6$PS$_5$Cl & Li$_6$PS$_5$Cl & Li$_6$PS$_5$Cl & Li$_6$PS$_5$Br & Li$_6$PS$_5$I \\
\hline \noalign{\vspace{2pt}}
Training anion disorder (\%) & 50 & 50 &  50 &  50 &   50\\
Extrapolating anion disorder (\%) & 25 &  75 &  100   & 100   & 100  \\
Energy RMSE (meV\,atom$^{-1}$) & 7.18 &  8.45 & 10.06  & 27.71  &  15.71 \\
Force RMSE (eV\,\AA$^{-1}$) & 0.168 &  0.181  &  0.175 &  0.200   &  0.213 \\

\end{tabular}
\end{ruledtabular}

\end{suptable*}

\newpage
\begin{supfigure*}
    \capstart
    \centering
    \includegraphics{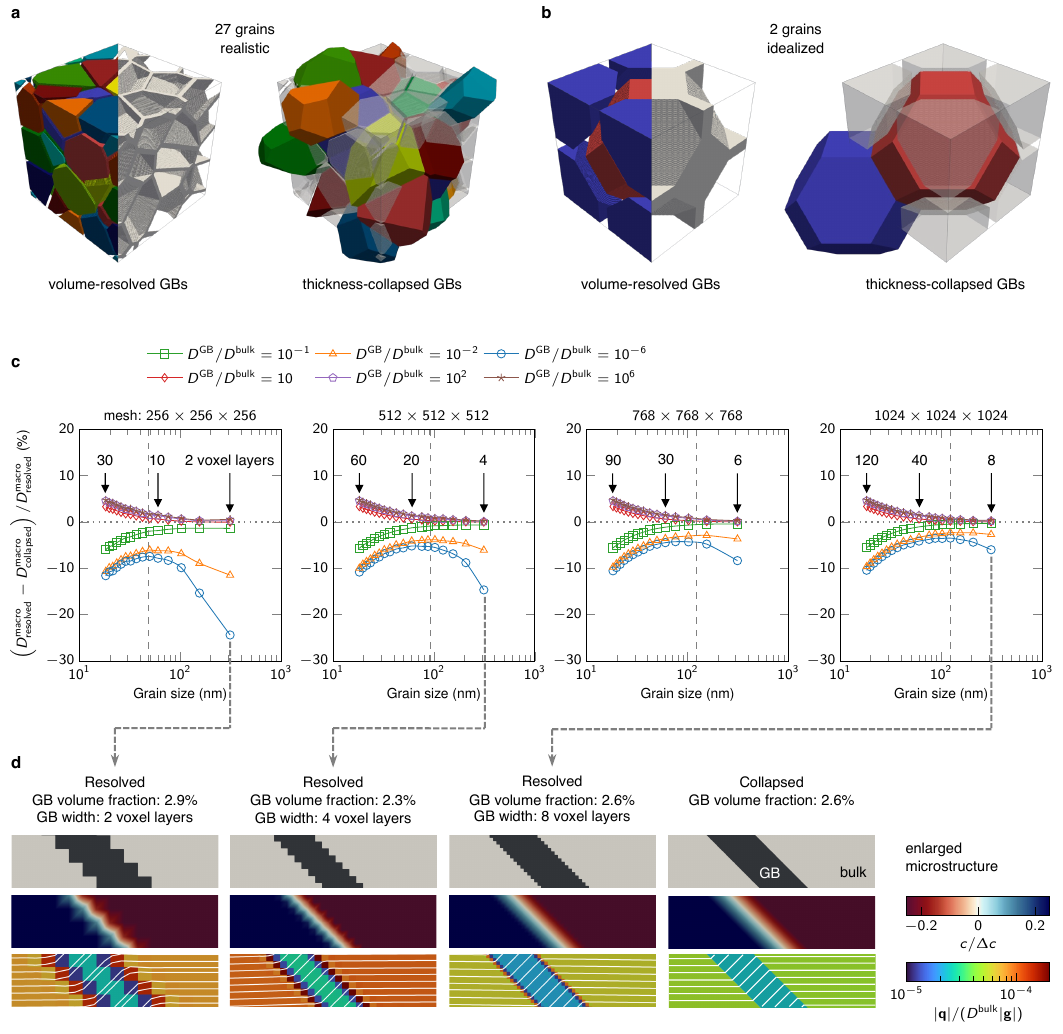}
    \caption[Finite element simulations with GBs]{
    \textbf{Finite element models and simulations for predicting macroscopic diffusivity in polycrystalline solid electrolytes.}
    \textbf{a}, Polycrystalline models with 27 grains used as a realistic volume element for finite element (FE) simulations in production runs. Voxel-based image with grain boundaries (GBs) of finite thickness (volume-resolved approach) and the crinkle-cut geometry featuring zero-thickness GBs (thickness-collapsed approach) are shown. 
    \textbf{b},~Idealized, isotropic polycrystalline models with two grains used to test the consistency of the two approaches. 
    %
    \textbf{c}, Consistency test of the two approaches using the idealized model with the same parametric setup. 
    Deviation in macroscopic diffusivities resulting from the two approaches depends on the mesh resolution of the volume-resolved model, average grain size, and the GB-to-bulk diffusivity ratio. 
    The GB width was varied to enable different grain sizes.
    Numbers at the arrows indicate the voxel layers used to resolve GBs. 
    The dashed line marks the grain size with the smallest deviation between the two approaches, converging to $\sim$100~nm at higher mesh resolutions.
    \textbf{d}, Enlarged view of the polycrystalline microstructure, the corresponding normalized concentration field $c$ and two-dimensional projection of the normalized flux field $\mathbf{q}$ from FE simulations using the two approaches. 
    The imposed concentration gradient $\mathbf{g}$ and the corresponding concentration jump $\Delta c$ were used for normalization.
    The ratio $D^{\mathrm{GB}}/D^{\mathrm{bulk}}=10^{-6}$ and grain size $\sim$300~nm are focused on. 
    Increasing mesh resolution reduces differences in concentration and flux fields between the two approaches, consistent with the deviation values.
    }
    \label{fig:femmodel}
\end{supfigure*}

\clearpage
\begin{suptable}[!ht]
    \capstart
    \caption[Examples of calculated GB volume fractions]{
    \textbf{Calculated grain boundary volume fraction for an illustrative two-grain polycrystalline model with cubic geometry.} 
    The simulation cell has a side length of 10~nm. The Voronoi seeds are located at fractional coordinates (0.7120, 0.4287, 0.1273) and (0.3314, 0.5767, 0.3072).
    }
    \label{tab:gb_ratio}
    \vspace{0.1cm}
    \sffamily
    \sansmath
    \selectfont
\begin{minipage}{0.55\linewidth}

\begin{ruledtabular}
\begin{tabular}{cc}
grain boundary width (nm) & grain boundary volume fraction (\%) \\
\hline \noalign{\vspace{2pt}}
0.3 & 10.614639521 \\
0.4 & 14.063393325 \\
0.5 & 17.338106036 \\ 
0.6 & 20.579787344 \\
0.7 & 23.666596413 \\ 
0.8 & 26.748743653 \\
0.9 & 29.758719355 \\
\rowcolor{gray!15}1.0 & 32.612385601 \\
1.1 & 35.470832884 \\
1.2 & 38.198748976 \\
1.3 & 40.934503078 \\
1.4 & 43.533254415 \\
\rowcolor{gray!15}1.5 & 46.082925051 \\
1.6 & 48.511920869 \\
1.7 & 50.918011367 \\
1.8 & 53.265763819 \\
1.9 & 55.476129055 \\
2.0 & 57.686384767 \\
2.1 & 59.776703268 \\
2.2 & 61.869255453 \\
2.3 & 63.853708655 \\
2.4 & 65.781814605 \\
\rowcolor{gray!15}2.5 & 67.630871385 \\
2.6 & 69.417306781 \\
2.7 & 71.171044558 \\
2.8 & 72.814197838 \\
2.9 & 74.456302822 \\
3.0 & 76.009458303 \\
3.1 & 77.504047751 \\
3.2 & 78.923345357 \\
3.3 & 80.303093791 \\
3.4 & 81.643644720 \\
3.5 & 82.880389690 \\
3.6 & 84.111364931 \\
3.7 & 85.245972872 \\
3.8 & 86.375099421 \\
3.9 & 87.435894459 \\
4.0 & 88.439571857 \\
4.1 & 89.386612922 \\
4.2 & 90.291553736 \\
4.3 & 91.164941341 \\ 
4.4 & 91.953337938 \\
4.5 & 92.732436955 \\
4.6 & 93.431632966 \\
4.7 & 94.121170789 \\
4.8 & 94.758135080 \\
4.9 & 95.344003290 \\
\end{tabular}
\end{ruledtabular}

\end{minipage}
\end{suptable}

\clearpage
\begin{supfigure*}[!ht]
    \capstart
    \centering
    \includegraphics{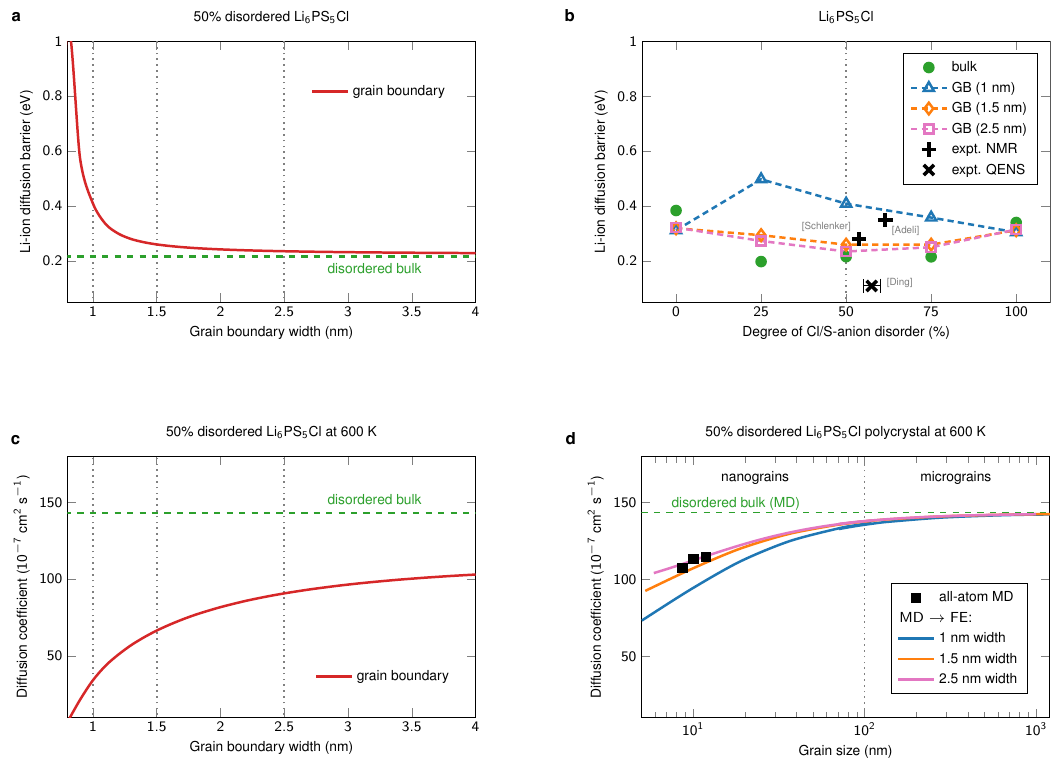}
    \caption[Sensitivity of diffusivity to GB width]{
    \textbf{Sensitivity analysis of grain boundary (GB) width.}
    A GB width of \qty{2.5}{nm} was derived from atomistic GB structures and was adopted for the main results in this study. 
    Results obtained using smaller widths of \qty{1.5}{} and \qty{1}{nm} are shown in the present figure for comparison.
    \textbf{a}, Predicted Li-ion diffusion barriers in grain boundaries estimated using different GB widths, exemplified for Li$_6$PS$_5$Cl with 50\% anion disorder.
    \textbf{b}, Li-ion diffusion barriers in Li$_6$PS$_5$Cl as a function of anion disorder.
    GB widths of \qty{1.5}{nm} and \qty{2.5}{nm} yield similar results and are consistent with experimental values with uncertainties (\citet{Adeli2019Jun, Schlenker2020Oct,Ding2025Jan}), whereas a width of \qty{1}{nm} leads to overestimated barriers.
    \textbf{c}, GB diffusivities at \qty{600}{\kelvin} estimated using different GB widths.
    \textbf{d}, Macroscopic diffusivities of 50\% anion-disordered Li$_6$PS$_5$Cl polycrystals at \qty{600}{\kelvin}.
    The use of smaller GB widths of \qty{1.5}{} and \qty{1}{nm} leads to inconsistencies between all-atom atomistic and continuum-scale predictions.
    }
    \label{fig:gbwidth}
\end{supfigure*}

\newpage
\begin{supfigure*}[!ht]
    \centering
    \includegraphics{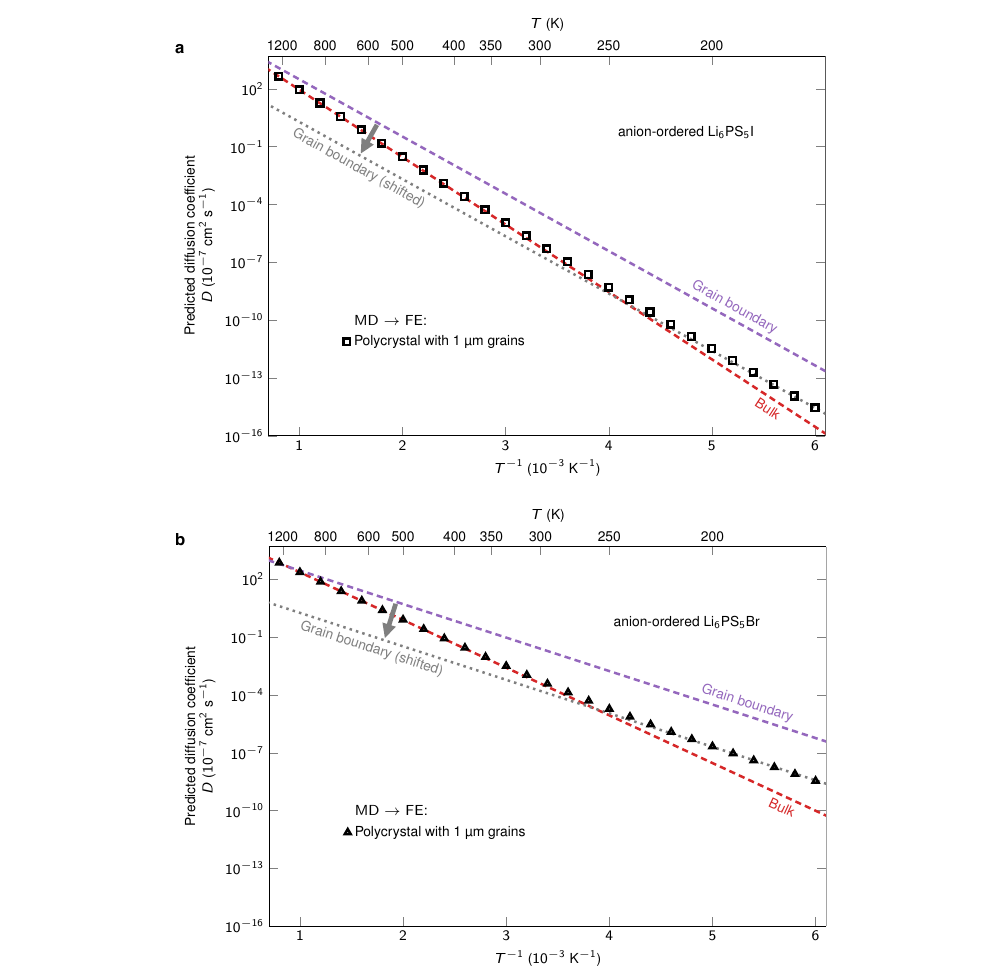}
    \caption[Non-Arrhenius in ordered argyrodites]{
    \textbf{Non-Arrhenius diffusion behavior arising from grain boundaries in anion-ordered argyrodites.}
    Temperature-dependent diffusivities of, \textbf{a}, anion-ordered Li$_6$PS$_5$I and, \textbf{b}, anion-ordered Li$_6$PS$_5$Br.
    Polycrystalline models with \qty{1}{\micro\meter}-sized grains are used for demonstration.
    Dashed lines are obtained from Arrhenius fits to bulk and grain-boundary diffusivities. 
    Additional dotted lines, shifted from the grain-boundary Arrhenius fits, highlight the grain-boundary-dominated regime at low temperatures.
    The non-Arrhenius behavior depends on the ratio of bulk to grain-boundary diffusivities and its temperature dependence.
    }
    \label{fig:non-arrhenius}
\end{supfigure*}

\newpage
\begin{supfigure*}[!ht]
    \centering
    \includegraphics{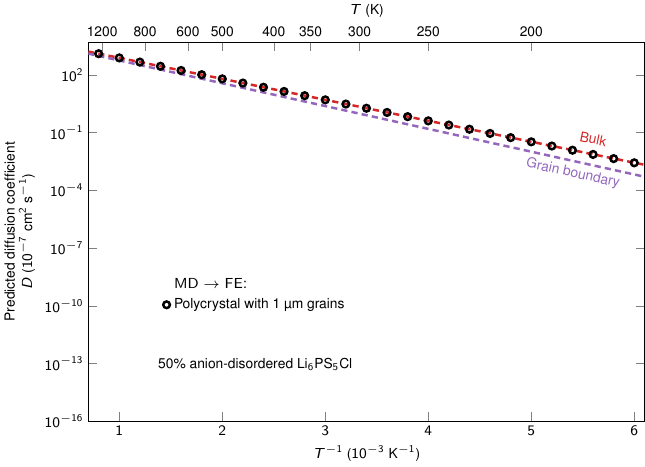}
    \caption[Arrhenius-like in disordered argyrodites]{
    \textbf{Diffusion behavior of polycrystalline anion-disordered argyrodites.}
    Results from 50\% anion-disordered Li$_6$PS$_5$Cl polycrystalline model are exemplified.
    Dashed lines are obtained from Arrhenius fits to bulk and grain-boundary diffusivities.
    Because bulk and grain-boundary diffusivities are comparable, the influence of grain boundaries is minimal, and the polycrystalline diffusivity closely follows the bulk Arrhenius relation. 
    Non-Arrhenius behavior is not clearly observed within the investigated temperature range.
    }
    \label{fig:dis_arrhenius}
\end{supfigure*}

\newpage
\begin{supfigure*}[!ht]
    \capstart
    \centering
    \includegraphics{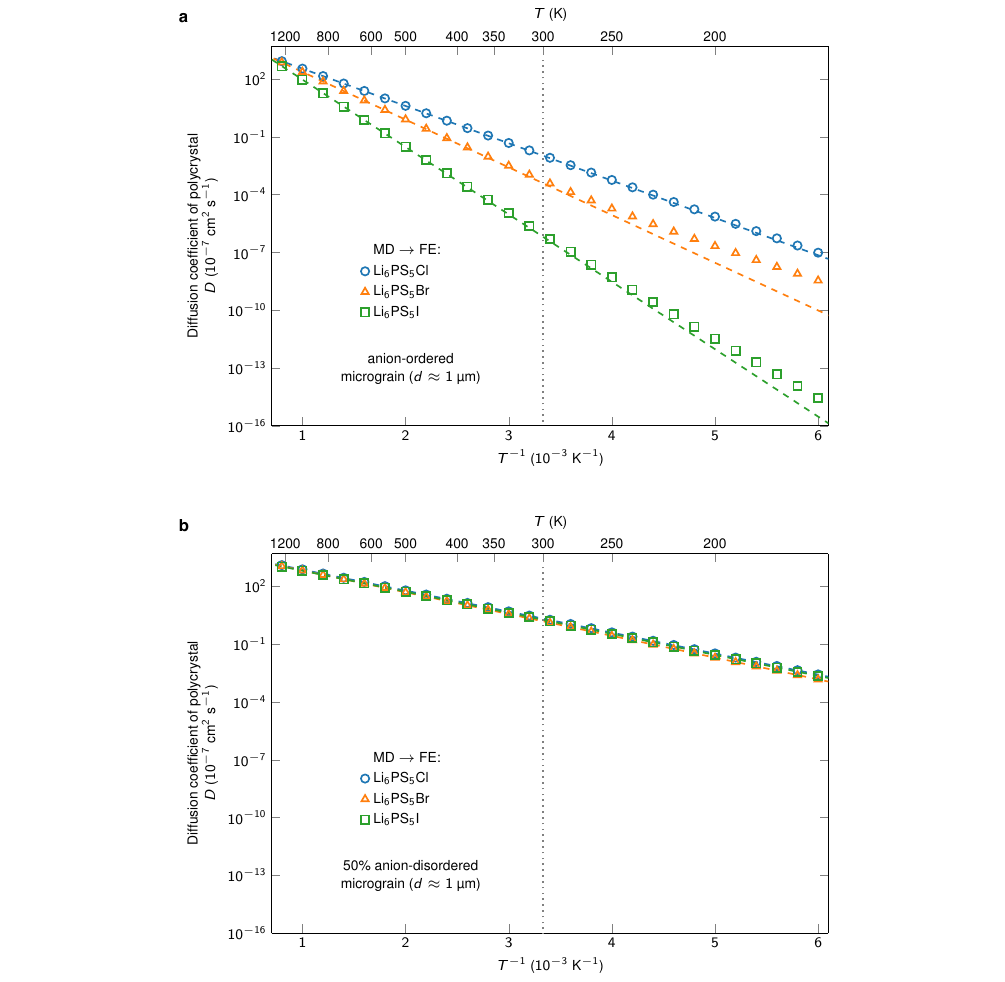}
    \caption[Diffusivities of argyrodites with \qty{1}{\micro\meter} grains]{
    \textbf{Comparison of predicted temperature-dependent diffusivities of polycrystalline Li\textsubscript{6}PS\textsubscript{5}\textit{X} with \textit{X} $\in$ \{\textrm{Cl}, \textrm{Br}, \textrm{I}\}.}
    \textbf{a}, Without anion disorder. 
    The non-Arrhenius diffusion behavior is observed for  Li$_6$PS$_5$Br and Li$_6$PS$_5$I, but cannot be clearly observed for Li$_6$PS$_5$Cl in the presented temperature range. 
    \textbf{b}, With anion disorder. 
    The non-Arrhenius diffusion behavior cannot be clearly observed for 50\% disordered Li$_6$PS$_5X$ with $X \in \{\textrm{Cl}, \textrm{Br}, \textrm{I}\}$ in the presented temperature range. 
    }
    \label{fig:ordervsdisorder}
\end{supfigure*}

\clearpage
\begin{suptable*}
    \capstart
    \caption[Study timeline of solid electrolyte GBs]{
    \textbf{Timeline of computational approaches for investigating Li-ion diffusion in inorganic solid electrolytes with grain boundaries (GBs).} 
    Molecular dynamics simulations with interatomic potentials were performed in these studies. 
    Yes or - indicates whether the corresponding model was considered in the given study. 
    Since 2022, machine-learning potentials have gained increasing popularity.
    } 
        \vspace{0.1cm}
	\begin{ruledtabular}
        {\sffamily
        \sansmath
        \selectfont
	\begin{tabular}{cccccc}
		Year  &  Interatomic potential{\normalfont\unsansmath\footnote{
        The pairwise potentials include the Morse, the Buckingham, and the Lennard--Jones forms, some combined with a Coulombic term. 
        The angular potential includes the Stillinger--Weber potential. 
        Machine-learning potentials, such as the neural network potential (NNP)~\cite{Singraber2019May,Singraber2019Mar}, the moment tensor potential (MTP)~\cite{Shapeev2016Sep, Novikov2020Dec, Podryabinkin2023Aug}, and the deep potential (DP)~\cite{Wang2018Jul} are highlighted.}}  & Solid electrolyte{\normalfont\unsansmath\footnote{
        The sulfide-based solid electrolytes are highlighted. }} & Single-type GBs{\normalfont\unsansmath\footnote{
        The notation with $\Sigma$ indicates the degree of misorientation of the GBs~\cite{Grimmer1974Mar}.}} & Polycrystal{\normalfont\unsansmath\footnote{
        Random GBs and triple junction structures are included in the polycrystalline simulation model.}} & Reference \\
        \hline \noalign{\vspace{2pt}}
            2017 & pairwise & Li$_7$La$_3$Zr$_2$O$_{12}$ & $\Sigma3$, $\Sigma5$ & - & \cite{Yu2017Nov} \\
            2018 & pairwise & Li$_3$OCl & $\Sigma3$, $\Sigma5$ & - & \cite{Dawson2018Jan} \\
            2018 & pairwise & Li$_7$La$_3$Zr$_2$O$_{12}$ & $\Sigma3$, $\Sigma5$, $\Sigma7$, $\Sigma9$, $\Sigma11$ & - & \cite{Shiiba2018Sep} \\
            2018 & pairwise & Li$_7$La$_3$Zr$_2$O$_{12}$ & $\Sigma5$ & - & \cite{Yu2018Nov} \\
            2021 & pairwise & Li$_{0.16}$La$_{0.62}$TiO$_3$ & $\Sigma2$, $\Sigma3$, $\Sigma5$ & - & \cite{Symington2021} \\
            2021 & pairwise + angular & LiZr$_2$(PO$_4$)$_3$ & - & yes & \cite{Kobayashi2022Mar} \\
            2021 & pairwise + angular & LiZr$_2$(PO$_4$)$_3$ & $\Sigma3$, $\Sigma5$, $\Sigma7$ & - & \cite{Nakano2021Nov} \\
            2021 & pairwise & Li$_{6.25}$La$_{3}$Zr$_{2}$O$_{12}$ & - & partially{\normalfont\unsansmath\footnote{
            Diffusivities of amorphous phases from atomic simulations were used for GBs in a phase field model to simulate polycrystals.}} & \cite{Heo2021Dec} \\
            [2.5pt]
            2022 & pairwise & \cellcolor{gray!15}Li$_{10}$GeP$_2$S$_{12}$ & - & yes & \cite{Dawson2022Feb} \\
            2022 & \cellcolor{gray!15}NNP & Li$_7$La$_3$Zr$_2$O$_{12}$ & $\Sigma5$ & - & \cite{Kim2022Jun} \\
            2022 & pairwise & Li$_3$OCl, Li$_3$OBr & - & yes & \cite{VanDuong2022} \\
            2023 & \cellcolor{gray!15}MTP & Li$_{0.375}$Sr$_{0.4375}$Ta$_{0.75}$Zr$_{0.25}$O$_{3}$ & $\Sigma3$, $\Sigma5$, $\Sigma51$ & - & \cite{Lee2023Apr} \\
            2023 & \cellcolor{gray!15}MTP & \cellcolor{gray!15}$\beta$-Li$_3$PS$_4$ & $\Sigma3$, $\Sigma5$, $\Sigma13$, $\Sigma25$ & - & \cite{Jalem2023Dec} \\
            2024  & \cellcolor{gray!15}MTP & \cellcolor{gray!15}Li$_6$PS$_5$Cl  & $\Sigma5$ &   -  & \cite{Kim2024Jun} \\
            2024  &  pairwise & Li$_{6.75}$La$_{3}$Zr$_{1.75}$Nb$_{0.25}$O$_{12}$  & $\Sigma3$, $\Sigma5$, $\Sigma7$, $\Sigma9$, $\Sigma11$  &  -  & \cite{Shiiba2024Jul} \\ 
            2024  &  pairwise & Li$_{2}$Ti$_{6}$O$_{13}$, Li$_{2}$Zr$_{6}$O$_{13}$  & -  &  yes  & \cite{Zulueta2024} \\ 
            2024  &  pairwise & Li$_{0.3125}$La$_{0.5625}$TiO$_{3}$  & $\Sigma5$  &  -  & \cite{Madrid2024Oct} \\ 
            2024  &  \cellcolor{gray!15}MTP & \cellcolor{gray!15}Li$_6$PS$_5$Cl  & $\Sigma3$, $\Sigma5$ &   -  & \cite{Ou2024Nov} \\ 
            2025  &  ReaxFF{\normalfont\unsansmath\footnote{
            The reactive force field (ReaxFF) developed in Ref.~\cite{Shin2018Aug} was used.}} & Li$_{1.3}$Al$_{0.3}$Ti$_{1.7}$(PO$_4$)$_3$  & - &   yes  & \cite{Ghosh2025Feb} \\ 
            2025  &  \cellcolor{gray!15}DP & LiCl{\normalfont\unsansmath\footnote{
            LiCl forms at the solid electrolyte interphase of chloride-based solid electrolytes, such as Li$_6$PS$_5$Cl~\cite{Kim2025Jun,Li2020Dec}.}}  & - &   yes  & \cite{Kim2025Jun} \\ [2pt]
            \hline \noalign{\vspace{2pt}}
            2026  &  \cellcolor{gray!15}MTP & \cellcolor{gray!15}Li$_6$PS$_5$Cl,\, Li$_6$PS$_5$Br,\, Li$_6$PS$_5$I  & $\Sigma3$,\, $\cdots$,\, $\Sigma51$,\, $\cdots$,\, $\mathbf{\Sigma99}${\normalfont\unsansmath\footnote{
            To our knowledge, $\Sigma 99$ is the single-type GB with the smallest misorientation angle (\SI{11.54}{\degree}) that has been investigated in solid electrolytes to date.}} &  yes  & this work \\
	\end{tabular}
        }
	\end{ruledtabular}
	\label{tab:history}
\end{suptable*}

\clearpage
\begin{supfigure*}
    \capstart
    \centering
    \includegraphics{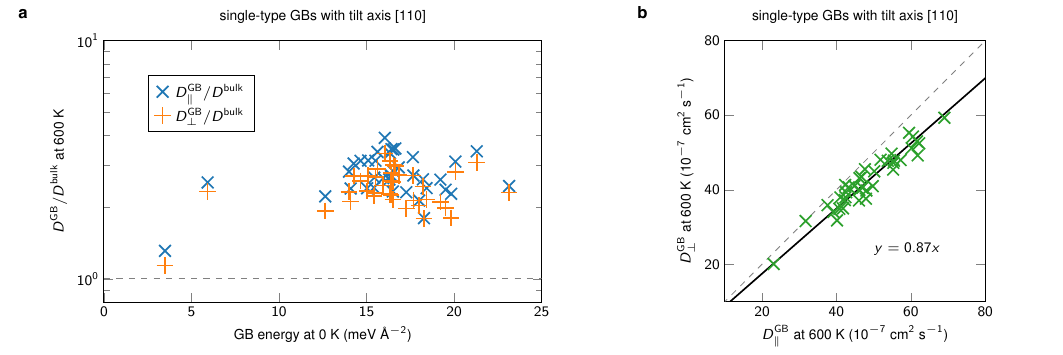}
    \caption[Li-ion diffusivities in single-type GBs]{
    \textbf{Analysis of Li-ion diffusivity in different single-type grain boundaries (GBs) of anion-ordered Li$_6$PS$_5$Cl.}
    Tilt GBs with a [110] rotation axis, ranging from $\Sigma3$ to $\Sigma99$ and covering misorientation angles from \SIrange{0}{180}{\degree}, are included. 
    \textbf{a}, Correlation between GB energy at \SI{0}{\kelvin} and GB to bulk diffusivity ratio at \SI{600}{\kelvin}. 
    GB diffusivities are decomposed into components parallel ($\parallel$) and perpendicular ($\perp$) to the GB plane. 
    The Borisov relation~\cite{Pelleg1966Sep} is not statistically supported by the present data. 
    \textbf{b}, Correlation between Li-ion diffusivity parallel and perpendicular to the GB plane. 
    GBs in Li$_6$PS$_5$Cl show weak anisotropy in diffusivity, with the perpendicular component reaching $\sim$87\% of the parallel value at 600~K. 
    }
    \label{fig:singlegb}
\end{supfigure*}

\clearpage
\begin{supfigure*}
    \capstart
    \centering
    \includegraphics{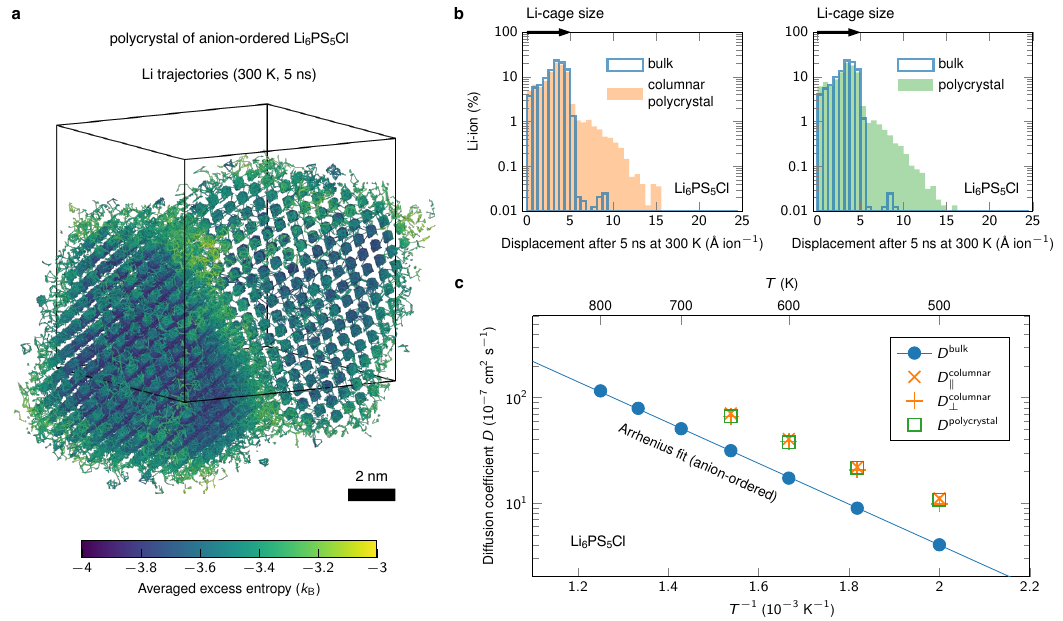}
    \caption[Columnar~vs.~polycrystalline models]{
    \textbf{Comparison of Li-ion diffusivity in anion-ordered Li$_6$PS$_5$Cl between columnar and polycrystalline models.}
    \textbf{a}, Trajectories of Li ions in a polycrystalline model after molecular dynamics simulations at \SI{300}{\kelvin} for \SI{5}{\nano\second}. Grain-boundary regions enclosing the grains exhibit elevated excess entropy, reflecting structural disorder. 
    \textbf{b}, Comparison of Li-ion displacement distributions in bulk, columnar, and polycrystalline models after \SI{5}{\nano\second} of molecular dynamics simulations at \SI{300}{\kelvin}. 
    The Li cage is \SI{\sim5}{\angstrom} in size. 
    Similar enhancement of inter-cage diffusion, indicated by Li-ion displacements exceeding \SI{10}{\angstrom}, is observed in both columnar and polycrystalline models. 
    \textbf{c}, Comparison of Li-ion diffusion coefficients calculated for all Li atoms in bulk, columnar, and polycrystalline models. 
    The number of Li atoms is comparable in these models, and Li ions exhibit similar diffusion coefficients in columnar and polycrystalline models.
    In the columnar model, Li-ion diffusion parallel ($\parallel$) to the grain-boundary plane is comparable to diffusion perpendicular ($\perp$) to it. 
    }
    \label{fig:colvspol}
\end{supfigure*}

\newpage
\begin{supfigure*}
    \capstart
    \centering
    \includegraphics{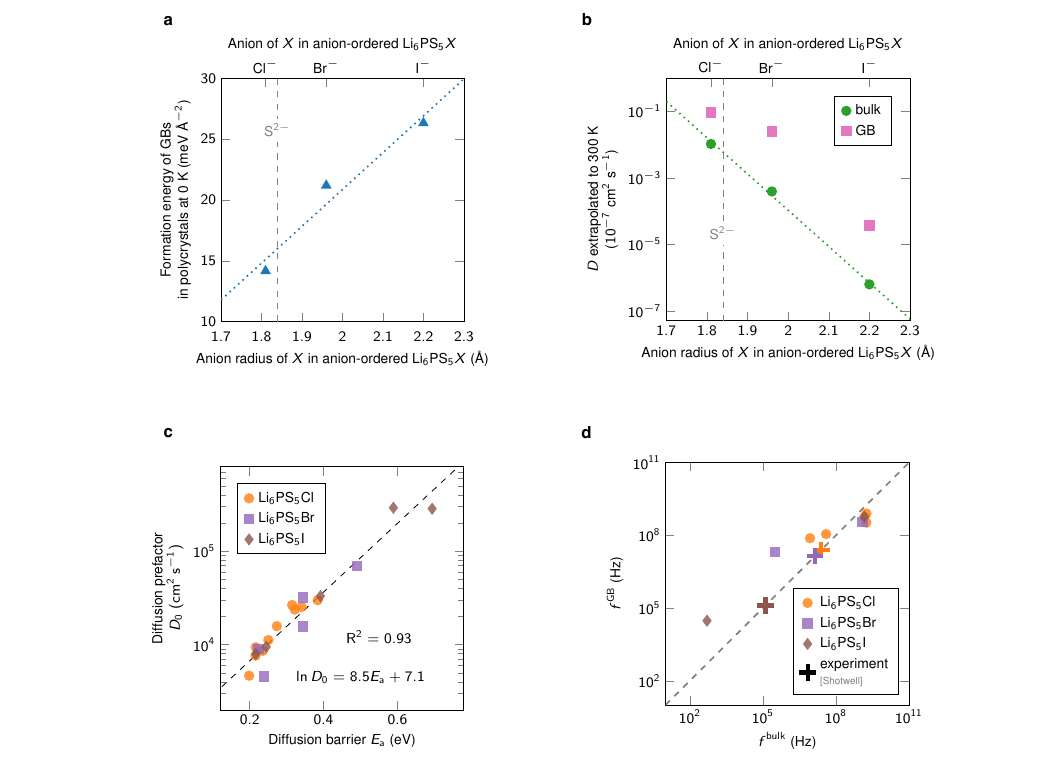}
    \caption[GB effects in argyrodite electrolytes]{
    \textbf{Grain-boundary (GB) effects in argyrodite solid electrolytes, Li\textsubscript{6}PS\textsubscript{5}\textit{X} with \textit{X} $\in$ \{\textrm{Cl}, \textrm{Br}, \textrm{I}\}.}
    \textbf{a}, Impact of the halide anion on GB formation energy at \SI{0}{\kelvin}. 
    The GBs are embedded in polycrystalline models with random grain orientations. 
    The ionic radius of S$^{2-}$ is similar to that of Cl$^-$. 
    GB formation energy positively correlates with the halide anion radius in argyrodites. 
    \textbf{b}, Predicted diffusion coefficients extrapolated to \SI{300}{\kelvin} based on Arrhenius fitting for bulk and GBs in argyrodites. 
    The GB values (averaged over all GBs in polycrystal) were extracted from polycrystalline models. 
    Diffusivity decreases with increasing anion radius of the halide species. 
    The GB-induced cage-opening effect enhances Li-ion diffusivity in all investigated argyrodites. 
    \textbf{c}, Correlation between the diffusion prefactor $D_{0}$ and diffusion barrier $E_{\mathrm{a}}$ obtained from Arrhenius fitting. 
    The Meyer--Neldel rule~\cite{Metselaar_Oversluizen_1984} is verified for argyrodites with structural (bulk and GB) and chemical (anion disorder) variations. 
    \textbf{d}, Correlation of characteristic Li-ion hopping frequencies $f$ between the bulk and GBs. Experimental data from \citet{shotwell_tetrahedral_2025}, which include contributions from both bulk and GBs, are shown assuming $f^{\mathrm{bulk}} = f^{\mathrm{GB}}$ for visual comparison.
    }
    \label{fig:chemspace2}
\end{supfigure*}

\newpage
\begin{supfigure}[!ht]
    \capstart
    \centering
    \includegraphics{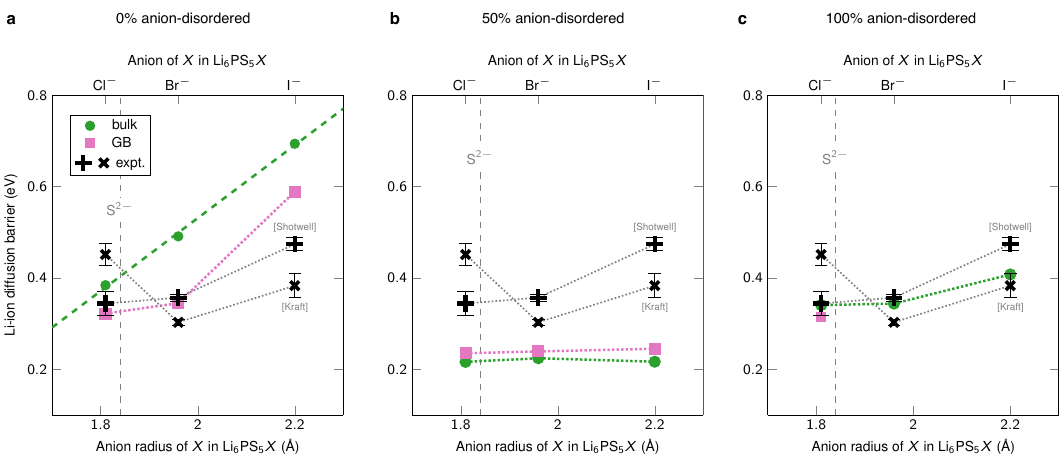}
    \caption[Li-ion diffusion barrier~vs.~anion radius]{
    \textbf{Influence of halide species on Li-ion diffusion barriers for argyrodites with different anion disorder.}
    Results are shown for Li$_6$PS$_5X$ with $X \in {\mathrm{Cl}, \mathrm{Br}, \mathrm{I}}$, with (\textbf{a}) 0\%, (\textbf{b}) 50\%, and (\textbf{c}) 100\% anion disorder.
    Consistency is observed across all argyrodites for the activation energy as a function of anion disorder.
    The bulk diffusion barriers decrease from 0\% to 50\% anion disorder and then increase from 50\% to 100\%.
    Grain boundaries hinder Li-ion diffusion at 50\% anion disorder, while they enhance diffusion at 0\%.
    The Li-ion diffusion barrier increases linearly with anion radius in anion-ordered bulk argyrodites, while exhibiting more complex behavior in other cases.
    Experimental values and uncertainties from electrochemical impedance spectroscopy measurements reported by \citet{shotwell_tetrahedral_2025} and \citet{Kraft2017Aug} are compared. 
    }
    \label{fig:disorder_trend}
\end{supfigure}

\clearpage
\begin{suptable*}[!ht]
    \capstart
    \caption[Overview of predicted diffusivities]{
    \textbf{Predicted Li-ion self-diffusivities and derived quantities of argyrodite solid electrolytes at 300~K.} 
    The anion disorder row indicates the degree of \textit{X}\!/S-anion disorder. 
    Grain boundary (GB) diffusion coefficients \smash{$D^{\mathrm{GB}}$} are from MD simulations of polycrystals, representing averages over all GB structures within them. 
    The GB-to-bulk ratio quantifies the impact of GBs on diffusion in the polycrystal with $\uparrow$ and $\downarrow$ for increase and decrease, respectively. 
    The $f$ rows show the predicted characteristic Li-ion hopping frequencies, which are consistent with experimental values from~\citet{shotwell_tetrahedral_2025}. 
    }
    \label{tab:allratio}
    \vspace{0.1cm}
    \sffamily
        \sansmath
        \selectfont
\begin{ruledtabular}
\begin{tabular}{lccccc}
 System & \multicolumn{5}{>{\columncolor{gray!15}}c}{Li$_6$PS$_5$Cl} \\
\hline \noalign{\vspace{2pt}}
Anion disorder (\%) & 0 &  25 & 50  & 75  & 100   \\
$D^{\mathrm{bulk}}$ ($10^{-7}$\,cm$^{2}$s$^{-1}$) & 0.010 & 2.153 & 2.184 & 1.808 & 0.048  \\
$D^{\mathrm{GB}}$ ($10^{-7}$\,cm$^{2}$s$^{-1}$) & 0.092 & 0.401 & 0.959 & 0.690 & 0.136 \\
$D^{\mathrm{GB}} / D^{\mathrm{bulk}}$ & 8.79 $\uparrow$ & 0.19 $\downarrow$ & 0.44 $\downarrow$ & 0.38 $\downarrow$ & 2.85 $\uparrow$  \\
$f^{\mathrm{bulk}}$ (Hz) & $8.61\times10^{6}$ & $1.77\times10^{9}$ & $1.80\times10^{9}$ & $1.49\times10^{9}$ & $3.92\times10^{7}$  \\
$f^{\mathrm{GB}}$ (Hz) & $7.57\times10^{7}$ & $3.30\times10^{8}$ & $7.90\times10^{8}$ & $5.68\times10^{8}$ & $1.12\times10^{8}$ \\ [3pt]
$f^{\mathrm{experiment}}$ (Hz) \tiny\textcolor{gray}{[Shotwell]} & \multicolumn{5}{>{\columncolor{gray!15}}c}{$2.50\times10^{7}$}  \\
\end{tabular}
\end{ruledtabular}

\vspace{15pt}

\begin{ruledtabular}
\begin{tabular}{lccccc}
 System &  \multicolumn{5}{>{\columncolor{gray!15}}c}{Li$_6$PS$_5$Br}  \\
\hline \noalign{\vspace{2pt}}
Anion disorder (\%) & 0 & & 50 &  & 100  \\
$D^{\mathrm{bulk}}$ ($10^{-7}$\,cm$^{2}$s$^{-1}$) & $3.9\times10^{-4}$ & & 1.522 & & 0.053 \\
$D^{\mathrm{GB}}$ ($10^{-7}$\,cm$^{2}$s$^{-1}$) & 0.025 & & 0.436 & & - \\
$D^{\mathrm{GB}} / D^{\mathrm{bulk}}$ & 64.06 $\uparrow$ & & 0.29 $\downarrow$  & &  - \\
$f^{\mathrm{bulk}}$ (Hz) & $3.19\times10^{5}$ & & $1.25\times10^{9}$ & & $4.38\times10^{7}$ \\
$f^{\mathrm{GB}}$ (Hz) & $2.04\times10^{7}$  &  & $3.57\times10^{8}$  & & - \\ [3pt]
$f^{\mathrm{experiment}}$ (Hz) \tiny\textcolor{gray}{[Shotwell]}  & \multicolumn{5}{>{\columncolor{gray!15}}c}{$1.37\times10^{7}$}  \\
\end{tabular}
	\end{ruledtabular}

\vspace{15pt}

\begin{ruledtabular}
\begin{tabular}{lccccc}
 System &  \multicolumn{5}{>{\columncolor{gray!15}}c}{Li$_6$PS$_5$I}    \\
\hline \noalign{\vspace{2pt}}
Anion disorder (\%)  & 0  &  &  50 &  & 100 \\
$D^{\mathrm{bulk}}$ ($10^{-7}$\,cm$^{2}$s$^{-1}$)  & $6.2\times10^{-7}$ & & 1.817 & & $4.1\times10^{-3}$\\
$D^{\mathrm{GB}}$ ($10^{-7}$\,cm$^{2}$s$^{-1}$) & $3.7\times10^{-5}$  & & 0.724 & & - \\
$D^{\mathrm{GB}} / D^{\mathrm{bulk}}$ &  59.25 $\uparrow$  & &  0.40 $\downarrow$ & & -\\
$f^{\mathrm{bulk}}$ (Hz)  & $5.05\times10^{2}$ & & $1.48\times10^{9}$ & & $3.31\times10^{6}$ \\
$f^{\mathrm{GB}}$ (Hz) &  $2.99\times10^{4}$ & & $5.88\times10^{8}$ & & -\\ [3pt]
$f^{\mathrm{experiment}}$ (Hz) \tiny\textcolor{gray}{[Shotwell]}  & \multicolumn{5}{>{\columncolor{gray!15}}c}{$1.27\times10^{5}$}  \\
\end{tabular}
	\end{ruledtabular}
\end{suptable*}

\newpage
\begin{supfigure*}
    \capstart
    \centering
    \includegraphics{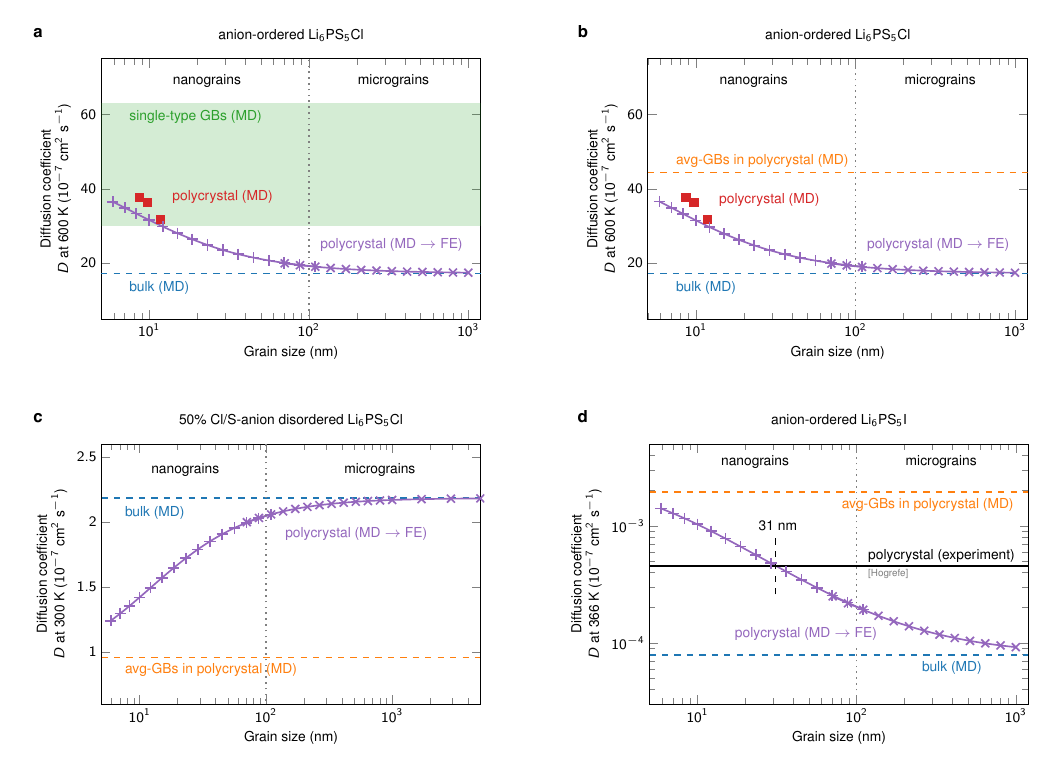}
    \caption[Predicted diffusivities~vs.~grain size]{
    \textbf{Predicted Li-ion diffusivities in polycrystalline argyrodite solid electrolytes vs.~grain size.} 
    \textbf{a}, Anisotropic diffusivities of randomly selected single-type grain boundaries (GBs) are used as input for finite element (FE) simulations. 
    Tests show that deviations arising from the choice of GB types are negligible at the macroscopic scale, owing to the small differences in diffusivity among single-type GBs. 
    Total diffusivities obtained from polycrystalline atomistic models demonstrate the consistency between atomistic and continuum simulations. 
    \textbf{b}, GB diffusivity extracted from polycrystalline models (avg-GBs in polycrystal) is used as input for FE simulations. 
    The macroscopic diffusivities from MD and from the MD $\rightarrow$ FE simulations for anion-ordered Li$_6$PS$_5$Cl shown in \textbf{a} and \textbf{b} are in excellent agreement, confirming that using avg-GBs values as inputs for FE simulations is appropriate. 
    \textbf{c}, Macroscopic diffusivity of Li$_6$PS$_5$Cl with 50\% Cl/S-anion disorder at \SI{300}{\kelvin}. 
    Diffusivity of the bulk structure is larger than that of GBs. 
    Large changes in diffusivity are predicted for polycrystals with nanograins ($d < 100$ nm). 
    \textbf{d}, Macroscopic diffusivity of anion-ordered Li$_6$PS$_5$I at \SI{366}{\kelvin}. 
    Experimental diffusivity can be reproduced with a polycrystalline microstructure having an average grain size of about \SI{31}{\nano\meter}. 
    }
    \label{fig:grainsize}
\end{supfigure*}

\newpage
\begin{supfigure*}
    \capstart
    \centering
    \includegraphics{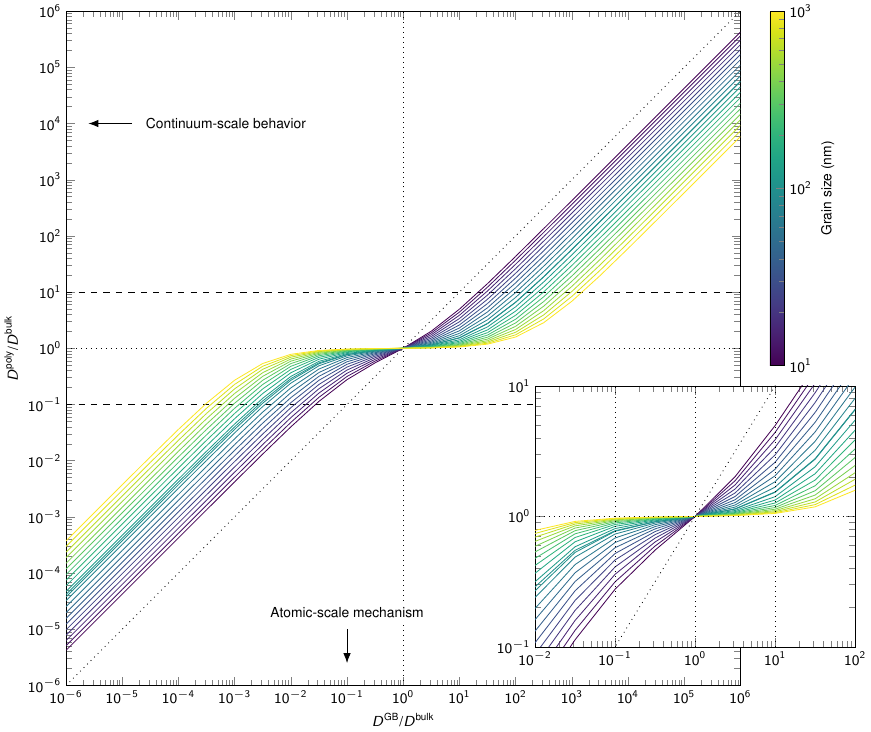}
    \caption[Map of atomistic- to continuum-scales]{
    \textbf{Precomputed map linking atomic-scale mechanisms to continuum-scale behavior in general polycrystalline solid electrolytes.}
    In a polycrystalline model with isotropic grain boundaries (GBs), the $x$-axis indicates the assumed GB diffusivity relative to the bulk, while the $y$-axis shows the resulting macroscopic diffusivity of the polycrystal relative to the bulk. 
    The line color represents the average grain size of the polycrystalline model, with the GB width of \SI{2.5}{nm}. 
    Dashed lines indicate a one-order-of-magnitude change in macroscopic diffusivity. 
    The inset shows an enlarged section of the precomputed map. 
    }
    \label{fig:ratio}
\end{supfigure*}

\newpage
\begin{supfigure*}
    \capstart
    \centering
    \includegraphics{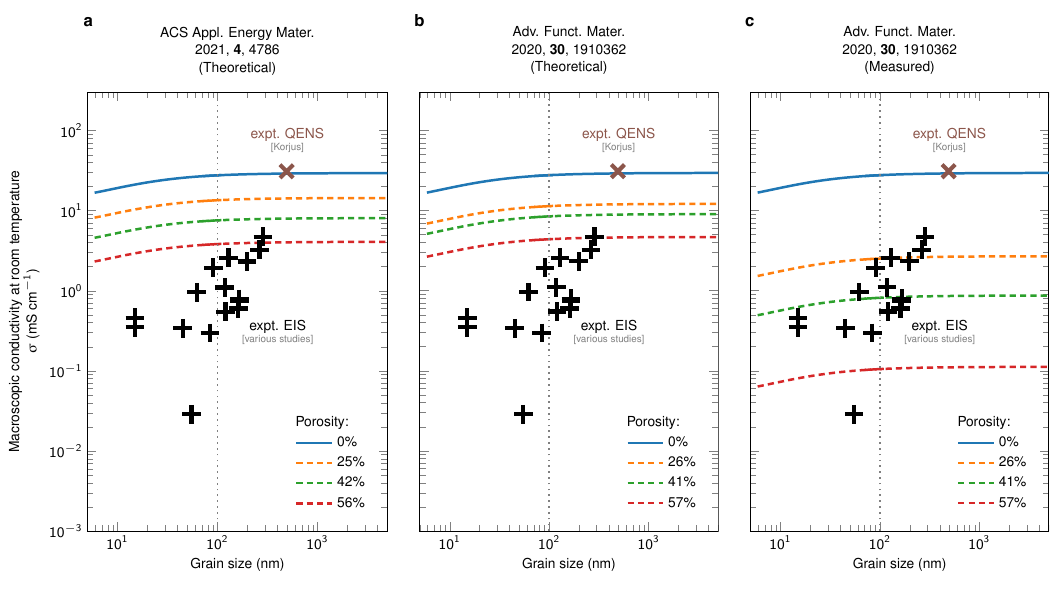}
    \caption[Diffusivities of porous argyrodites with GBs]{
    \textbf{Impact of porosity on macroscopic conductivity of solid electrolytes at 300~K.}
    Li$_6$PS$_5$Cl with 50\% anion disorder is considered.
    The grain-size effect is estimated for a close-contact polycrystal using our multiscale modeling framework, while the porosity effect is applied as a shift on top of these close-contact results, based on theoretical models (ACS~Appl.~Energy~Mater.~\cite{Neumann_Hamann_Danner_Hein_Becker-Steinberger_Wachsman_Latz_2021} and Adv.~Funct.~Mater.~\cite{ Hamann_Zhang_Gong_Godbey_Gritton_McOwen_Hitz_Wachsman_2020}) and experimental measurements~\cite{Hamann_Zhang_Gong_Godbey_Gritton_McOwen_Hitz_Wachsman_2020}.
    Accounting for porosity reduces the macroscopic conductivity by one to two orders of magnitude, bringing predictions for anion-disordered Li$_6$PS$_5$Cl closer to experimental observations. 
    }
    \label{fig:poriosity}
\end{supfigure*}

\clearpage

\makeatletter
\renewcommand\bibsection{%
  \section*{Supplementary references}%
}
\makeatother